\documentclass[preprint]{aastex}
\newcommand{\ul}[2]{\left(\frac{#1}{#2}\right)}
\newcommand{\ee}[1]{\times10^{#1}}

\slugcomment{Draft}

\shortauthors{Wojdowski, Clark, \& Kallman}
\shorttitle{An X-Ray Spectroscopic Study of the SMC~X-1/Sk~160 System}

\begin{document}

\title{An X-Ray Spectroscopic Study of the SMC~X-1/Sk~160 System}

\author{Patrick S. Wojdowski\altaffilmark{1} and George W. Clark}
\affil{Physics Department and Center for Space Research, Massachusetts
Institute of Technology, Cambridge, MA 02139}

\and

\author{Timothy R. Kallman}
\affil{NASA/Goddard Space Flight Center, Greenbelt, MD 20771}

\altaffiltext{1}{Current Address: Lawrence Livermore National
Laboratory, P. O. Box 808, L-41, Livermore, CA 94551, 
\email{patrickw@virgo.llnl.gov}}

\begin{abstract}
We have investigated the composition and distribution of the wind of
Sk 160, the supergiant companion of the X-ray star SMC~X-1, by
comparing an X-ray spectrum of the source, obtained with the {\it
ASCA\/} observatory, during an eclipse with the computed spectra of
reprocessed radiation from circumstellar matter with various density
distributions.  We show that the metal abundance in the wind of Sk~160
is no greater than a few tenths of solar, as has been determined for
other objects in the Magellanic Clouds.  We also show that the
observed X-ray spectrum is not consistent with the density
distributions of circumstellar matter of the spherically symmetric
form derived for line-driven winds, nor with the density distribution
derived from a hydrodynamic simulation of the X-ray perturbed and
line-driven wind by \citet{blo95}.
\end{abstract}


\section{INTRODUCTION} 

The evidence that some of the first discovered cosmic X-ray sources
were binary star systems included periodic ``low-states'' which were
recognized as eclipses by a companion star.  However, it was observed
that during these eclipses there was still a residual flux of X-rays.
\citet{sch72} observed a residual flux from Cen~X-3 during its
eclipses and proposed that it might be due to ``the slow radiative
cooling of the gas surrounding the system which is heated by the
pulsating source''.  In observations with the GSFC Cosmic X-ray
Spectroscopy experiment on board {\it OSO-8}, \citet{bec78} detected a
significant flux of residual X-rays during three eclipses of the
massive X-ray binary Vela~X-1 and argued -- on the grounds that the
residual eclipse X-ray had, within limits, the same spectrum as the
uneclipsed flux -- that this eclipse radiation was most likely due to
X-ray scattering around the primary star.

Observations of the P-Cygni profiles of ultraviolet lines in massive
stars showed that strong winds are a ubiquitous feature of massive
stars \citep{mor67b}.  These winds are driven as the UV radiation from
the star transfers its outward momentum to the wind in line
transitions \citep{luc70,cas75}.  In high-mass X-ray binaries (HMXBs)
then, an obvious source of circumstellar material exists to scatter or
reprocess the X-rays from the compact object and allow X-rays from
these sources to be observed during eclipse.

Information on the composition and dynamics of HMXB winds may provide
important clues for the understanding of binary evolution
(e.g. mass-loss rates).  Information on the metal abundances may also
inform the study of binary evolution.  In other galaxies, such as the
Magellanic Clouds, X-ray reprocessing in the winds of HMXB may provide
another way to measure the metal abundances in those galaxies and
therefore inform the study of the evolution of those galaxies.

Properties of HMXB winds have been inferred in studies which interpret
the the X-ray spectrum at various phases in terms of the absorption,
emission, and scattering by the wind material.  Studies of X-ray
absorption during eclipse transitions have shown that the atmospheric
densities near the surfaces of the massive companions decrease
exponentially atmospheres with scale heights $\sim$1/10 of the stellar
radius \citep{sch72,sat86,cla88}.  Such exponential regions are not
predicted by the theory of \citet{cas75} or subsequent refinements
\citep[e.g.][]{pau86,kud89} which assume that the wind is spherically
symmetric. X-rays photo-ionize the wind, destroying the ions
responsible for the radiation driving, and shutting off the wind
acceleration, in the region illuminated by the X-ray star.
Hydrodynamic simulations have been done to explore the behavior of
winds in HMXB under the influrence of X-ray ionization and of X-ray
heating of the exposed face of the companion star
\citep{blo90,blo94,blo95}.

Several studies have compared observations with the results of Monte
Carlo calculations with model winds.  In these calculations,
individual photons were tracked though trial wind density
distributions where they could scatter from electrons or be absorbed.
Fluorescent photons were emitted following inner shell ionization and
these photons were tracked through the wind in the same way.
\citet{lew92} found that {\it Ginga} spectra of Vela~X-1 could be
reproduced with a density distribution of the form of a radiatively
driven wind with an exponential lower region.  \citet{cla94} attempted
to reproduce {\it Ginga} spectra of 4U 1538-52 with a similar density
distribution but with added components resembling structures in the
simulations of \citet{blo90}.  These authors were able to reproduce
the observed spectrum above 4.5 keV. They attributed an excess at
lower energies in part to a dust-scattered halo.  \citet{woo95a} found
that {\it Ginga} spectra of SMC~X-1 could be reproduced from the
calculated density distribution of \citet{blo95}.

Spectra from the moderate-resolution ($E/\Delta E \sim 15$ at 1 keV)
X-ray CCDs on {\it ASCA} --- and the prospect of high resolution X-ray
spectra from {\it AXAF}, {\it XMM}, and {\it ASTRO-E} --- present new
opportunities for using X-ray spectra in the study of HMXB winds.  In
an observation of Vela~X-1 with {\it ASCA\/}, approximately 10
emission features emerged in the spectrum of the system viewed during
eclipse of the X-ray source when the much more intense, featureless
continuum radiation was occulted.  Most of these emission features
were identified as K$\alpha$ emission lines from helium and
hydrogen-like ions of astrophysically abundant metals \citep{nag94}.
The Monte Carlo spectral calculations described above did not include
recombination to highly ionized atoms which is responsible for these
lines and so could not have predicted this spectrum.
Moderate and high-resolution spectra provide a powerful diagnostic of
the conditions in the wind through study of this recombination
radiation. 

\citet{ebi96} have estimated the size of the ionized regions in the
wind in the Cen~X-3 system from the magnitude of the recombination
lines measured with {\it ASCA} and \citet{lie96} have estimated the
emission measure of ionized regions in the wind of the Cyg~X-3 system
from the recombination lines and narrow recombination continua in the
{\it ASCA} spectrum.  \cite{sak99} have estimated the differential
emission measure of the wind of Vela~X-1 from the recombination
features.  They also developed a method to compute the emission
spectrum of recombination from a given matter distribution and have
inferred the presence of additional dense matter in the Vela~X-1 wind
from the presence of fluorescence lines.

In this paper, we present a study of the circumstellar matter in the
SMC~X-1/Sk~160 binary system based on observations of SMC~X-1 in and
out of eclipse. Using the XSTAR program, we calculate the spectra of
reprocessed radiation for a range of values of the ionization
parameter ($\xi\equiv L/nr^2$).  We show that the eclipse spectrum of
SMC X-1, which is similar in shape to the out-of-eclipse spectrum,
resembles the spectra of X-rays reprocessed in material with either
low or high ionization parameters but not the spectra expected from
material with intermediate ionization parameters which should exhibit
strong recombination features.  We then describe a method to calculate
the spectra of reprocessed X-rays from an arbitrary wind distribution.
This method takes account of electron scattering and fluorescence and
calculates the reprocessed emisson by summing the diffuse emission and
absorbing along the lines of sight to the observer.  We use this
method to predict the spectrum of X-rays emitted by a wind with the
density distribution of \citet{blo95}.  Because our calculations
include recombination radiation and because the {\it ASCA} data is
sensitive to recombination spectral features, we obtain a conclusion
which contradicts that of \citet{woo95a}.  The matter distribution of
\citet{blo95} would produce strong recombination radiation features
which are excluded by the observed {\it ASCA} eclipse spectrum.

In Section~\ref{obs} we discuss our observations of SMC~X-1 with {\it
ASCA}.  In Section~\ref{repro}, we discuss our calculations of the
spectra of reprocessed X-rays from homogenous, optically thin gas, and
compare it to the {\it ASCA} data.  In Section~\ref{winds}, we
describe our procedure for computing the spectrum of reprocessing from
a 3-D distribution of gas and apply it to the density distribution
from the hydrodynamic simulation by \citet{blo95}.

\section{{\it ASCA\/} OBSERVATIONS}
\label{obs}

SMC~X-1 has been observed twice by the {\it ASCA\/} observatory: once,
while the source was uneclipsed, soon after launch in April of 1993
during the performance verification (PV) phase, and again in October
of 1995, while the source was eclipsed.  The out-of-eclipse
observation, yielded 17,769 seconds of SIS0 data and 17,573 seconds of
SIS1 data after screening according to the default criteria described
in the {\it ASCA} Data Reduction Guide.  During the eclipsed
observation, SMC~X-1 was observed over a period of approximately
94,000 seconds resulting in 32,838 seconds of SIS0 data and 32,239
seconds of SIS1 data after screening.

We used the FTOOLS package to extract light curves and spectra from
these data and the XSPEC \citep{arn96} program to do the spectral
analysis.  The source regions of the CCD chips were chosen to be
squares centered on the image of SMC~X-1 and $383\arcsec$ on a side.
The remainders of the chips were used as the background regions.  To
derive spectra and light curves, the number of counts in each spectral
or time bin in the background region was subtracted (after correcting
for the difference in the relative sizes of the source and background
regions) from the counts in the corresponding source bin.

Light curves for both observations are plotted in Figure~\ref{lcs}
along with a light curve from Ginga \citep[from an observation
described by][]{woo95a} in the energy band where that instrument's
energy band overlaps {\it ASCA\/}'s.  The {\it Ginga\/} light curve
illustrates the ``high state'' behavior of SMC~X-1 with sharp eclipse
transitions at phases $\pm$0.07.  A spectrum was extracted from the
out-of-eclipse {\it ASCA\/} observation using all but a few hundred
seconds at the end where the count rate started to decrease, possibly
due to an onset of the low state (not visible in Figure~\ref{lcs}).
The resulting exposures were 17,226 seconds for SIS0 and 17,026
seconds for SIS1.  Spectra were extracted from the entire time of the
eclipse observation though that observation extends beyond the time of
nominal eclipse.  The fact that the flux from SMC~X-1 differed by no
more than a factor of 3 between times inside and outside of the
nominal eclipses indicates that the system was in the low-state ---
with the compact object blocked by the precessing accretion disk
\citep{woj98}.  Since the X-rays detected during that observation have
been reprocessed in the stellar wind, all of the data from that
observation, including that from outside of the nominal eclipse, were
used in our analysis in order to improve the signal-to-noise ratio of
the eclipse spectrum.  To further improve the signal the energy
channels, which oversample the detector resolution, were grouped so
that each channel has at least 50 counts.

Due to the high count rate in the out-of-eclipse observation, the
errors due to counting statistics were small compared to the
systematic errors due to the instrument calibration and the data from
the two detectors could not simultaneously be fitted to a single trial
spectrum.  For this reason, only the data from the SIS0 detector,
which may be better calibrated, were used in spectral fits for this
observation.  The spectrum derived from the out-of-eclipse observation
was well fit by a model which consists of a power-law plus two broad
gaussian components (one near 0.9 keV and one near 6 keV) absorbed by
a small column of cold interstellar material.  The fitted model for
the photon flux is
\begin{equation}
{\cal F}(E) = e^{-\sigma(E)N_{\rm H}} [f_{\rm pl}(E)+f_{\rm
ga1}(E)+f_{\rm ga2}(E)]
\times\left\{ \begin{array}{ll}
1                                                   & E\leq10{\rm keV} \\
\exp\left(-\frac{E-10{\rm keV}}{15{\rm keV}}\right) & E>10{\rm keV}
\end{array}\right.
\label{spec_exp}
\end{equation}
where
\begin{eqnarray}
f_{\rm pl}(E)=K_{\rm pl}(E/1keV)^{-\alpha}, \\
f_{{\rm ga}i}(E)=\frac{K_{{\rm ga}i}}{\sigma_i \sqrt{2\pi}}
\exp\frac{-(E-E_i)^2}{2\sigma_i^2},
\end{eqnarray}
and $\sigma(E)$ is the cross-section of interstellar absorption of
\citet{mor83}.  While the sensitivity of the {\it ASCA\/} detectors
does not extend to the high energy cut-off, it has been measured with
{\it Ginga} \citep{woo95a} and is included here for consistency with
the XSTAR calculations.  The observed spectra and best fit model are
plotted in Figure~\ref{unespec}.  Though they were not used in the
spectral fits, the data from the SIS1 detector are included in this
plot.  There is a discrepancy between the two detectors near 1.3 keV
which is comparable to the channel-to-channel variations in
observations of the supernova remnant 3C 273 \citep{orr98} which were
used for calibration.  The best fit values of the parameters for this
and the two fits described below are tabulated in
Table~\ref{specpars}.  Approximately the same result was obtained by
\citet{sta97} for this data set.  The fitted value for the column of
neutral hydrogen lies between values estimated by interpolation from
neighboring directions in 21-cm emission surveys: $4.6\times 10^{20}$
cm$^{-2}$ for a galactic survey with $1^\circ$ resolution
\citep{dic90} and $4.5\times 10^{21}$ cm$^{-2}$ for a survey of the
SMC with $98\arcsec$ resolution \citep{sta99}.  The flux of this model
spectrum in the 13.6 eV--13.6 keV band (the band in which luminosity
is defined in XSTAR) is $6.44\times 10^{-10}$ erg~cm$^{-2}$~s$^{-1}$.
For isotropic emission, this corresponds to a luminosity of $1.9\times
10^{38}$ erg~s$^{-1}$ for a distance of 50~kpc.  A source with this
spectrum would give an on-axis count rate of 1.1 count s$^{-1}$
\footnote{This count rate was computed by comparing the model flux to
that of the Crab \citep{too74} in each of the 2--3, 3--5, and 5--10
keV bands and assuming that the count rate of the Crab is 25 count
s$^{-1}$ in each of these bands.}  in the {\it RXTE} All-Sky Monitor
whereas the observed count rate of SMC~X-1 in the All-Sky Monitor, at
the peak of the high state is approximately 3 count s$^{-1}$
\citep{woj98} which implies that the luminosity of SMC~X-1 in its high
state is approximately $5\times 10^{38}$ erg s$^{-1}$.

In the eclipse observation, the systematic calibration errors are much
smaller than the statistical errors due to the low count rate.
Therefore, spectral models were fit simultaneously to both the SIS0
and the SIS1 detectors.  To fit the spectrum derived from the
observation during eclipse the same model that fit the out-of-eclipse
data, scaled down in intensity, was tried.  This is the spectrum that
would be expected if the intrinsic spectrum of the accreting neutron
star did not change between the two observations and the eclipse
spectrum were due only to Compton scattering of the X-rays from the
source.  Relative to the rescaled out-of-eclipse spectrum, the eclipse
data show a large excess at energies greater than 4 keV and a small,
narrow excess near 1.8 keV, the approximate location of the
fluorescence line of neutral silicon (Figure~\ref{eclscal}).  An
acceptable fit for the eclipse data can be obtained by allowing the
parameters of the broad 6 keV feature to vary from their values scaled
from the out-of-eclipse spectrum.

A separate fit to the data points in the 1.5--2.0 keV range was done
using a power law continuum and a narrow emission feature.  The energy
of this emission feature was found to be $1.775\pm0.020$ keV and the
flux was $1.5\pm0.5\times 10^{-5}$ photon$\,$s$^{-1}$cm$^{-2}$.  This
feature is near the location of the detector's internal fluoresence
peak and so we must consider the possibility that this feature is
spurious.  For the {\it AXAF} ACIS CCDs, which are similar to the {\it
ASCA\/} CCDs, the probability that an X-ray with energy greater than
the Si K edge will cause fluorescence in the detector is 1\% just
above the K edge and decreases to 0.2\% at 4 keV \citep{pri98}.  The
SIS0 screened event list has 4769 photons with energies above the
silicon K edge which should result in less than $\sim$50 fluorescent
photons produced in the detector.  The measured line flux ($1.5\times
10^{-5}$ photon$\,$s$^{-1}$cm$^{-2}$) corresponds to 74 photons in
this spectrum.  A miscalibration of the probability of fluorescence
events of order 100\% would be necessary to result in a spurious
detection of this magnitude.  Furthermore, in the data from the
out-of-eclipse observation, in which the source has nearly the same
spectrum, there is no deviation from the smooth spectral model at this
energy.  Also, the measured energy of this feature differs from the
1.74 keV of neutral silicon that would come from the detector but is
consistent with the energy of the fluorescent line of partially
ionized silicon. Still, confirmation of this feature will require
better observation.

We also tried a reflection model (the ``hrefl'' model by T. Yaqoob in
XSPEC) to fit the $\sim$6 keV feature --- i.e. we assumed that the
source spectrum could be described by a power law plus a broad $\sim$
0.9 keV component, and that the observed spectrum includes a component
due to reflection of that spectrum from cold, neutral gas.  For the
out-of-eclipse spectrum, we assumed that the ``escape fraction'' was
unity --- i.e. the neutron star was directly visible.  For the the
eclipsed spectrum we assumed the same model normalization --- i.e. the
neutron star had the same spectrum and luminosity as during the
out-of-eclipse observation.  The best fit parameters for fits to the
reflection models are given in Table~\ref{reffit}.  The out-of-eclipse
spectrum requires a covering fraction which is greater than unity and
therefore unphysical.

\section{Spectral Models of Reprocessed Radiation}
\label{repro}

For the densities lower than approximately 10$^{12}$~cm$^{-3}$, with
which we are concerned, the heating, cooling, ionization, and
recombination are dominated by interactions between single photons
from a point source and photons and gas particles
(e.g. photo-ionization) and by two-particle interactions
(e.g. recombination).  If the radiation from the point source is not
significantly attenuated, the rate of photon-particle interactions is
proportional to $Ln/r^2$ and the rate of two-particle interactions is
proportional to $n^2$.  Here, $L$ is the luminosity of the point
source, $r$ is the distance from it, and $n$ is the density of
hydrogen (neutral and ionized. 
Spontaneous de-excitations (single particle
transitions) happen on timescales short compared to the time between
two-body interactions and can therefore be considered not as
independant processes, but as part of the two-body interaction which
produces the excited state. Therefore, for a given gas composition and
a given radiation spectrum, the state of the gas is a function of the
ratio of the quantities $Ln/r^2$ and $n^2$: the ionization parameter,
$\xi\equiv L/nr^2$, defined by \citet{tar69a}. 

Reprocessed radiation includes re-emission from recombination,
bremsstrahlung, and fluorescence as well as electron scattering.
Bremsstrahlung and recombination radiation are two-body processes, so
for a given state of the gas, their intensity is proportional to
$n^2$.  Since, for a given gas composition and spectrum of incident
radiation, the state of the gas is a function only of $\xi$, the
volume emission coefficient for these processes may be written:
\begin{equation}
 j_\nu^{(1)}=f_\nu(\xi)n^2.
 \label{2body_jnu}
\end{equation}
Photo-ionization of inner shell electron may result in a raditive
cascade which fills the inner shell vacancy. Because the resultant
radiation is due to photo-ionization, its contribution to the volume
emission coefficient is proportional to $Ln/r^2$ and can be written
as:
\begin{eqnarray}
 j_\nu^{(2)} & =g_\nu(\xi)Ln/r^2 \\
             & =\xi g_\nu(\xi) n^2 .
 \label{ph-par_jnu}
\end{eqnarray}
The contribution of electron scattering, which we consider to be an
absorption followed by emission, like fluorescence, is proportional to
$Ln/r^2=\xi n^2$.  Thus, for a given spectrum of primary radiation and
gas composition, the entire spectrum of reprocessed emission, from
given volume element $dV$ is a function of the ionization parameter
scaled by $n^2dV$.

We used the XSTAR program \citep[v1.43,][]{kal99} to calculate the
spectrum of the X-rays that would be emitted by gas illuminated by
radiation with the spectrum of SMC~X-1.  XSTAR assumes a point
radiation source at the center of a spherically symmetric nebula and
calculates the state of the gas and the continuum and line radiation
emitted by the gas throughout the nebula.  We took the spectrum of the
central point source to be that derived from the out-of-eclipse
observation (Equation~\ref{spec_exp}) but with the absorbing column
set to zero.  We set the density of the gas at a constant value of
$10^{-3}$ cm$^{-3}$ and did calculations for the metal abundances
equal to the solar value \citep{and89} and less by 0.5, 1, 1.5, and
2.0 dex.  We chose a low gas density so that all optical depths would
be negligible.  These calculations are similar to Model 1 of
\citet{kal82}.

While XSTAR includes some effects of electron scattering, it does not
explicitly compute the ``emission'' due to electron scattering.
Therefore we added this component to the emissivities computed with
XSTAR.  Photons much lower in energy than $m_e c^2=$511 keV scatter
from free electrons with little change in frequency with the Thompson
cross-section.  Photons may also scatter from bound electrons as if
the electrons were free if the binding energies are much less than the
photon energies.  In gases of astrophysical interest approximately
98\% of the electrons are contributed by hydrogen and helium.  Since
the greatest binding energy in either of these two elements is 54.4
eV, for purposes of Compton scattering of X-rays of greater than
approximately 0.5 keV it is a very good approximation to assume that
the electron density is a constant fraction of the hydrogen
density. For the solar abundance of helium relative to hydrogen there
are approximately 1.15 electrons per hydrogen atom.  Thus, for X-rays
in the energy range 0.5--50 keV, the spectrum of Compton scattered
radiation is identical to the spectrum of input radiation and does not
depend on the ionization state or temperature of the plasma.  We
therefore added to the volume emissivity, the quantity
\begin{equation}
j_{\nu{\rm ,scattered}}= 1.15\times \frac{L_\nu}{L}
\frac{\sigma_T}{4\pi}\xi n^2
\end{equation}

The total spectrum of reprocessed X-rays from gas with 1/10 solar
metal abundance for several ionization parameters is shown in Figure
\ref{spec_zeta}.  These spectra modified by $7\ee{20}$ cm$^{-2}$ of
interstellar absorption and convolved with the {\it ASCA\/} response
matrix are shown in Figure~\ref{folded_spec_zeta}.  In the range
$1\lesssim \log\xi \lesssim 3$, the spectrum contains many strong
features from recombination to the K shells of astrophysically
abundant elements.  At lower values of the ionization parameter,
astrophysically abundant elements are not ionized up to the K shell
and therefore produce spectral features in the X-ray band only by
photo-ionization of K-shell electrons and subsequent radiative
cascades.  At higher values of the ionization parameter, the abundant
elements approach complete ionization and the electron temperature
approaches a limit.  Therefore, the emissivity due to recombination
approaches a limit while the continuum emissivity due to Compton
scattering continues to increase linearly with $\xi$.  The absorption
cross-section is plotted in Figure~\ref{opac}.

\subsection{Single $\xi$ Spectra and the {\it ASCA\/} Eclipse Spectrum}
\label{single_xi}
The {\it ASCA\/} spectrum of SMC~X-1 in eclipse was compared to the
convolved spectra of reprocessed radiation over a range of abundances
and ionization parameters. None of the reprocessing mechanisms
discussed above can reproduce the observed excess in the eclipse
spectrum around 6 keV.  Therefore we considered only the data for
energies less than 4 keV. The eclipse spectra can be fit by
reprocessing models with $\log\xi\gtrsim 3$ or by models with
$\log\xi\lesssim 1$ and metal abundance less than approximately 1/10
of solar (Figure~\ref{cont_z_xi}).  In the range $1<\log\xi<3$, even
for a metal abundance as low as 1/100 of solar, the calculations
predict a flux below 1 keV from recombination features that is
inconsistent with the observed spectrum
(Figure~\ref{spec_zeta_data_n_mo}b).  For $\log\xi>3$ recombination is
very weak relative to Compton scattering so the spectrum is
insensitive to metal abundance and satisfactory fits may be obtained
for abundance as large as solar (Figure~\ref{spec_zeta_data_n_mo}a).
The fluorescent features present in the spectral models for
$\log\xi<1$ are not as strong relative to the recombination features
as in the models for intermediate ionization.  This allows the data to
be fit with $\log\xi<1$ as long as the metal abundance relative to
solar is less than approximately 0.2
(Figure~\ref{spec_zeta_data_n_mo}c).  In contrast to the $\log\xi>3$
regime however, the spectrum of reprocessed X-rays contains
fluorescent emission lines.

The emission line detected at $1.775\pm0.020$ keV may be the
fluorescent K$\alpha$ line of \ion{Si}{2}
\footnote{We describe fluorescence lines according to the charge state
of the atom at the time the line is emitted, after photo-ionization of
the inner shell electron has occurred --- e.g. K shell
photo-ionization of neutral iron results in the emission of a Fe II
K$\alpha$ photon.}  or some combination of this line and the
fluorecent K$\alpha$ lines of \ion{Si}{3}--V all of which have
K$\alpha$ lines near 1.74 keV.  It may also be due, at least
partially, to a higher ionization stage, such as \ion{Si}{9}, which
has a fluorescent K$\alpha$ line at 1.77 keV. The feature cannot be
due to recombination to helium-like silicon (which would produce an
emission line at 1.84 keV) because under conditions necessary to
produce that line, recombination radiation from oxygen and other
elements would produce a large flux below 1 keV which is not seen.
The presence of this fluorescence line and the lack of strong
recombination features indicate that a significant fraction of the
emission comes from gas with $\log\xi<1$.  The reprocessing spectral
model with $\log\xi=0$ and metal abundance equal to one-tenth of solar
has a single emission line at 1.740 keV.  With the model normalization
fit to the SMC~X-1 elipse spectrum, this line has a flux of
$5.2\ee{-6}$ photon s$^{-1}$cm$^{-2}$.  The flux of the observed
feature is $1.5\pm0.5\ee{-5}$ photon s$^{-1}$cm$^{-2}$.  If this line
flux is correct, it indicates that the silicon abundance in SMC~X-1 is
at least two tenths of solar.

Except in the range $1<\log\xi<3$, the flux from the reprocessing
models is dominated by Compton scattering.  Therefore, the emission
measure ($n^2V$) necessary to reproduce a given luminosity with a
single $\xi$ reprocessing model should be proportional to the inverse
of the ionization parameter.  To confirm this, the best-fit value of
$\chi^2$ was computed on a grid of values of $\log\xi$ and the
normalization parameter ($K\equiv(n^2V/4\pi D^2)\times 10^{-14}{\rm
cm^5}$).  Contours from these fits are plotted in
Figure~\ref{cont_xi_norm}.  Indeed, the best fits are in a narrow
region of parameter space around the line defined by ($K\xi=1.7$).

To determine what amount of material with $1<\log\xi<3$ is allowed by
the observed spectra, fits were done to model spectra consisting of
reprocessing from gas at two ionization parameters.  Both components
had metal abundances fixed at one tenth of solar.  The first component
had an ionization parameter fixed at the the lowest calculated value
($\log\xi=-2.95$).  The ionization parameter and the normalization of
the second component were stepped through a grid of values and the
normalization of the first component was varied to minimize $\chi^2$.
Contours of the minimized $\Delta\chi^2$ relative to the best fit are
plotted in Figure~\ref{cont_xi_norm_2comp}.  As noted above, model
spectra for $\log\xi>3$ and for $\log\xi<1$ have the same shape since
the emission is dominated by Compton scattering.  For these values of
the ionization parameter, good fits can be obtained for normalizations
up to a value such that $K\xi\sim1.7$ as above.  For $1<\log\xi<3$
however, smaller normalizations are allowed.  For $\log\xi=2$,
$K\lesssim 3\ee{-4}$.  For a distance of 50 kpc, this corresponds to
an emission measure $n^2V \lesssim 9\ee{57}$ cm$^{-3}$.

\section{Spectra From Reprocessing in Model Winds}
\label{winds}

\subsection{Spectral simulation algorithm}
\label{algorithm}

We devised a procedure to calculate the spectrum of X-rays from a
central point souce reprocessed in a 3-D matter distribution.  The
flux received by an observer at a distance $d$ from a diffuse source
can be written schematically as
\begin{equation}
{\cal F}_\nu= \frac{1}{4\pi d^2}\int j_\nu({\bf
x})e^{-\tau_\nu({\bf x})}dV 
\end{equation}
where ${\bf x}$ is the spatial vector, $j_\nu$ is the volume emission
coefficient --- the energy output 
of the gas per unit volume, frequency, time, and solid angle in the
direction of the observer.  The optical depth $\tau_\nu({\bf x})$
between the point of emission and the observer.  The optical depth is
calculated
\begin{equation}
\tau_\nu({\bf x})=\int_{\bf x}^{{\bf x}_{\rm observer}}
	\sigma_\nu({\bf x})n({\bf x})dx
\end{equation}
where $\sigma_\nu$, is the cross-section for absorption and scattering
out of the line of sight per hydrogen atom.  Again, we consider a
photon received by the observer as having been ``emitted'' at the
place where it last interacted with the gas.

We map the density distribution onto a rectilinear grid such that
lines of sight to the observer are parallel to one of the axes. Then,
for every grid cell, we calculate the ionization parameter.  Then,
starting at grid cells opposite the observer, we look up the spectrum
of reprocessed emission for that ionization parameter, scale by
$n^2V$, and then add that to a running total emission spectrum.  At
the next grid cell we attenuate the emission spectrum from cells
behind according to the cross-sections for that ionization parameter
scaled by $nl$ where $l$ is the length of the grid cell, and so on to
compute the total spectrum of emission in the direction of the
observer.  Emission is assumed to come only from gas which is
illuminated by the point radiation source.  Absorption in
unilluminated material is taken to be equal to absorption from gas
with the lowest ionization parameter in the XSTAR table.  Emission
from points of gas which are not visible to the observer (i.e.  those
behind the companion star) are not included in the summation.

The flux received by the observer is equal to this luminosity divided
by 4$\pi$ times the distance squared.  The emission spectrum is
thereby converted to a flux and is output to a FITS format XSPEC Table
Model \citep{arn95} which is easily imported to the XSPEC spectral
fitting program \citep{arn96}.  and can easily be convolved with
instrument response matrices for comparison with observed spectral
data.

Our assumption that the direct radiation from the compact object is
not significantly attenuated in the gas may cause an overestimate of
the radiation from circumstellar material with substantial optical
depth.  The cross-sections plotted in Figure~\ref{opac} can be as
large as $10^{-21}{\rm cm^2}$ per hydrogen atom.  Therefore, this
algorithm will begin to fail for column densities of order
$10^{21}{\rm cm^{-2}}$ or greater.  The fact that this algorithm does
include absorption of the reprocessed radiation on its way from the
reprocessing cites to the observer does compensate for this error to
some extent however.  The only reprocessed radiation that will be seen
from a region of large optical depth is from the part of its surface
which is both exposed to the radiation source and visible to the
observer.  The algorithm would not be accurate if the radiation source
was surrounded by a small shell of optically thick gas.  In this case,
the algorithm would calculate emission from distant gas which was
actually shadowed by the dense shell.  However, if the optically thick
material does not subtend a large solid angle about the radiation
source, the error due to shadowing should be small.  In regions where
the density is so high that single grid cells are optically thick, the
neglect of absorption by material within the same cell in which it was
produced will cause the emission from those optically thick cells to
be overestimated by a factor approximately equal to the optical depth
of the cells.  If the optically thick material subtends only a small
solid angle at the radiation source and is not optically thick on
length scales much less than one pixel, the algorithm will calculate
the spectrum accurately except that the small portion of the emission
which is from the dense material will be overestimated by a factor of
no more than a few.

\subsection{ Hydrodynamic Simulation}

The spectrum of reprocessed emission from the \citet{blo95}
hydrodynamic simulation wind was synthesized using the algorithm
described in Section~\ref{algorithm}.  The density distribution on the
spherical grid from the hydrodynamic simulation was interpolated onto
a rectilinear grid with similar resolution: 50 grid points along the
radius of the simulation, equal to $1.43\ee{11}$ cm per grid point.
No points in the hydrodynamic simulation had densities no larger than
approximately $3\ee{11}{\rm cm^{-3}}$ so no pixel had an optical depth
significantly greater than one and the overestimate of the contribution
from optically thick cell was no greater than a factor of a few.
While the gas distribution does have regions of high density near the
radiation source, the high density material subtends a small angle in
the orbital plane and is mostly confined to the orbital plane so the
error in the simulated spectra due to this gas should not be large.
The spectral simulation was carried out for X-ray luminosities of 1,
1.7, 3, 6, 10, 17, and 30 times $10^{38}$ erg s$^{-1}$ for metal
abundances equal to solar and less by 0.5, 1.0, 1.5, and 2.0 dex. The
distribution of density and ionization parameter (for $L_X=3\ee{38}$
erg s$^{-1}$ for this model is shown in Figure~\ref{blo_pict}.

The SMC~X-1 eclipse spectrum was fit by interpolation on this grid of
models.  With the distance to SMC~X-1 fixed at 50 kpc, reasonable fits
are obtained for the luminosity in a narrow range around $6.4\ee{38}$
erg s$^{-1}$ and for abundances less than a few hundredths of solar
(Figure~\ref{cont_z_l_blo_fixd}).  Though a reasonable fit to the
global spectrum can be obtained (Figure~\ref{spec_blo_bf}), the lack
of a silicon line in the model spectrum indicates that the model is
deficient.  Furthermore, the best fit metal abundance is very low
compared to other measurements of the abundances in the SMC
\citep{wes97}.  Both the reason for the low abundance and for the lack
of the silicon feature can be seen in the differential emission
measure, plotted in Figure~\ref{dem_blo}.  For an X-ray luminosity of
$6\ee{38}$ erg s$^{-1}$, the hydrodynamic simulation contains gas with
$\log\xi>3$ and also some gas with $1<\log\xi<3$ but no gas with
$\log\xi<1$.  The presence of gas with $1<\log\xi<3$ produces strong
recombination emission features and only by setting the metal
abundances to be very low, can the calculated spectra be made to agree
with observed spectrum.  While the neglect in the algorithm of
absorption between the X-ray source and the reprocessing point may
have caused an overestimate of the recombination emission by a factor
of a few, the total emission measure of material with $2<\log\xi<3$ in
the \citet{blo95} model is $1.33\ee{59}$ cm$^{-3}$ compared to the
lower limit of $9\ee{57}$ cm$^{-3}$ for single components with
$\log\xi$ in that range derived in the previous section
(Figure~\ref{cont_xi_norm_2comp}).  The lack of a silicon emission
feature in the model can be explained by the absence of material at
low ionization.  The presence of the silicon fluorescence feature in
the observed spectrum indicates that there exists gas in the wind of
SMC~X-1 that is more dense than any of the gas in the hydrodynamic
simulations.  A hydrodynamic simulation with higher spatial resolution
might resolve the gas distribution into smaller, denser clumps and
move the peak of the emission measure distribution below $\log\xi=1$
where it would fluoresce but not emit recombination radiation in the
{\it ASCA\/} band.

\subsection{Absorption of the Direct Radiation}

We now explore the validity of our approximation that absorption of
radiation along lines of sight from the neutron star can be neglected
and that the spectrum of reprocessed radiation is a function only of
$\xi$ and spectrum of the radiation from the neutron star. Examination
of the \citet{blo95} density distribution --- plotted in
Figure~\ref{blo_pict} --- shows that the largest column densities
occur in dense clumps. Since the densest material has the lowest
ionization parameter and material at lower ionization has greater
opacity, these are the places where our approximation is most likely
to be invalid.  The contour denoting the highest density in
Figure~\ref{blo_pict} denotes a density of $10^{12}$~cm$^{-3}$.  The
distance from the neutron star to the first clump of this density is
that of thirteen grid cells which is equal to $1.86\ee{12}$ cm which
implies $\log\xi=2.26$.  We ran XSTAR with the density equal to
$10^{12}$~cm$^{-3}$, the luminosity equal to $6.4\ee{38}$ ergs
s$^{-1}$ and $\log\xi$ at the inner radius equal to 2.26.  In runs
with these parameters, ionization fronts like those in the optically
thick models of \citet{kal82} formed where the column depth reached
approximately $2\ee{23}$ cm$^{-2}$.  In Figure~\ref{col}, spectra of
reprocessed radiation are shown for a point before the ionization
front and after the ionization front --- with the reprocessing
spectral model from optically thin gas at the same ionization
parameter for comparison.  Before the ionization front, the spectra
are almost identical for the optically thin case and the optically
thick case for a given ionization parameter.  After the ionization
front, the spectrum of reprocessing is cut off below a few keV.
Therefore, for column depths less than $2\ee{23}$ cm$^{-2}$,
absorption along the paths from the radiation source to the
reprocessing sites can be ignored.  Of the paths which begin at the
neutron star, only those which go through the largest and densest
clumps have column densities of this magnitude so the error due to the
unaccounted for absorption is small.

\subsection{Spherically Symmetric Winds}
A spherically symmetric power-law wind distribution of the type
derived by \citet{cas75} for a radiation driven wind provides another
model to test against the observed spectrum.  In this type of wind,
the density is decribed by
\begin{equation}
n(r)=\frac{\dot{M}}{4\pi r^2 v_\infty \mu m_{\rm p}}(1-R_\star/r)^{-\beta}
\label{eqn:cak}
\end{equation}
where $R_\star$ is the radius of the star, $\dot{M}$ is the mass loss
rate, $m_{\rm p}$ is the proton mass, 
and $\mu$ is the mean molecular weight (the number of proton masses
per hydrogen atom in the gas, $\sim 1.34$).  In eclipse, very little
of the material near the stellar surface is both illuminated by the
X-ray source and visible.  Therefore the expression in parentheses in
Equation~\ref{eqn:cak} is near unity and the density is approximately
proportional to $r^{-2}$ in the region of interest.  If the binary
separation is not large compared to the size of the companion, then
the ionization parameter does not vary far from the value it would
have everywhere if the X-ray source were at the center:
\begin{equation}
\xi_0=4\pi v_\infty \mu m_{\rm p} L \dot{M}^{-1}
\label{eq:xi_0_approx}
\end{equation}
The density can then be written in terms of $\xi_0$.
\begin{equation}
n(r)=\frac{L}{\xi_0 r^2}
\label{eq:n_xi}
\end{equation}
The relation $K\xi=1.7$ derived in Section~\ref{single_xi} must hold
approximately for $\xi=\xi_0$.  In order to estimate $\xi_0$, we
estimate $K$ for a spherical wind.
\begin{equation}
K  =  \frac{1}{4\pi D^2}\int n^2 dV \ee{-14}
\end{equation}
and using Equation~\ref{eq:n_xi},
\begin{eqnarray}
K & = & 10^{-14} (4\pi D^2)^{-1} \int_{R_\ast}^\infty n^2 dV \nonumber
\\
\label{pl_eq}
 & = &10^{-14} (4\pi D^2)^{-1} \xi_0^{-2}L^2\int_{R_\ast}^\infty
r^{-4}4\pi r^2 dr \nonumber \\
 & = &10^{-14} D^{-2} \xi_0^{-2}L^2  {R_\ast}^{-1} 
\end{eqnarray}
Then, for $D$=50 kpc, $L=3\ee{38}$ erg s$^{-1}$, and $R_\ast=17
R_{\sun}$, $K\xi_0\approx1.7$ implies $\xi_0\approx2.4\ee{4}$ or less
if only part of the flux is due to the extended wind.  From
Equation~\ref{eq:xi_0_approx}, we find
\begin{equation}
 \xi_0 \approx 4.5\ee{3}\ul{L}{10^{38}{\rm erg\ s^{-1}}}
 \ul{\dot{M}}{10^{-6}M_{\sun}{\rm yr}^{-1}} \ul{v_\infty}{10^3{\rm km\
 s^{-1}}}
\end{equation}
So wind parameters of typical B-type stars ($10^{-7}$--$10^{-6}
\dot{M_{\sun}}{\rm yr}^{-1}$, $v_t\sim 1500{\rm km\ s^{-1}}$) result
a wind at high ionization and does not emit strong recombination
features. 

However, such a spherically symmetric high-ionization wind cannot
produce the observed silicon emission feature.  A component with
$\log\xi<1$ must be included.  An exponential region of the
atmosphere, of the type inferred from extended eclipse transitions,
may be a good candidate for such a low ionization region.  An
atmosphere with a scale height much less than the stellar radius would
be illuminated and visible in only a small region. This region has the
same distance from the compact object as does the center of the
companion star ($1.95\ee{12}$cm for SMC~X-1).  For this region to have
$\log\xi<1$, a density of approximately $10^{14}$ cm$^{-3}$ is
required for an X-ray luminosity of $5\ee{38}$ erg s$^{-1}$.
\citet{woo93} fit {\it Ginga} spectra during eclipse transitions to
absorption by wind models which had the form of Equation~\ref{eqn:cak}
but with an expontial region at the base of the wind.  At the minimum
radius of illumination and visibility, these models have densities no
greater than $5\ee{12}$ cm$^{-3}$. At the stellar surface these models
have densities no greater than $6\ee{13}$ cm$^{-3}$.  However, only
very near eclipse transitions can material at the stellar surface be
both illuminated by the X-ray source and visible to the observer.  The
distribution of $\log\xi$ for the best fit parameters to one of the
eclipse transitions is plotted in Figure~\ref{hyb_zeta} for an X-ray
luminosity of $3\ee{38}$ erg s$^{-1}$.

\section{Conclusions}

Through comparison of spectra from {\it ASCA} data from observations
of SMC~X-1 during an eclipse and archival data from outside of
eclipse, we derive the following conclusions:
\begin{enumerate}
\item The X-ray spectrum of SMC~X-1 has approximately the same form in
eclipse and out of eclipse.  This indicates that most of the X-ray
reprocessing in the circumstellar matter is due to Compton scattering.
\item The lack of strong recombination features in the eclipse
spectrum indicates that most of the gas in the wind of SMC~X-1 is
either highly ionized ($\log\xi>3$) or lowly ionized
($\log\xi<1$). Very little material is in intermediate ionization
states.
\item We find evidence of a small but significant emission feature
near the energy of the fluorescence line of neutral silicon (1.74
keV).  The presence of this feature, if confirmed by better
measurements, would indicate that a significant amount of gas must
have very low ionization.
\item The \citet{blo95} model of the density distribution derived by
3-D hydrodynamic simulation cannot reproduce the observed eclipse
spectrum.  It contains a large amount of material with $1<\log\xi<3$
which would produce a large flux of recombination radiation.  The
spectral resolution of {\it ASCA\/} makes these observations very
sensitive to these features.  However, no recombination features are
detected.  Also, this model does not predict any material at low
enough ionization to produce the observed silicon fluorescence line.
A hydrodynamic simulation with higher spatial resolution might resolve
smaller, denser clumps and produce an emission measure distribution
which would reproduce the observed eclipse spectrum.
\item A smooth, spherically symmetric wind could be sufficiently
ionized so as not to emit recombination radiation features which would
have been detectable in our observation.  However, it is difficult to
construct such a wind to be consistent with observation of X-ray
absorption in eclipse transitions, and with the observed intensity of
the silicon fluorescence line.
\end{enumerate}

\acknowledgements We thank J. Blondin and J. Woo for making available
to us the density distrubution from their hydrodynamic simulation in
machine readable format.  This work was supported in part by NASA
grant NAG5-2540.  PSW received support from a NASA Graduate Student
Researchers Program Fellowship through Goddard Space Flight Center
(NGT5-57). 

\bibliographystyle{apj} 
\bibliography{ms}

\clearpage

\figcaption[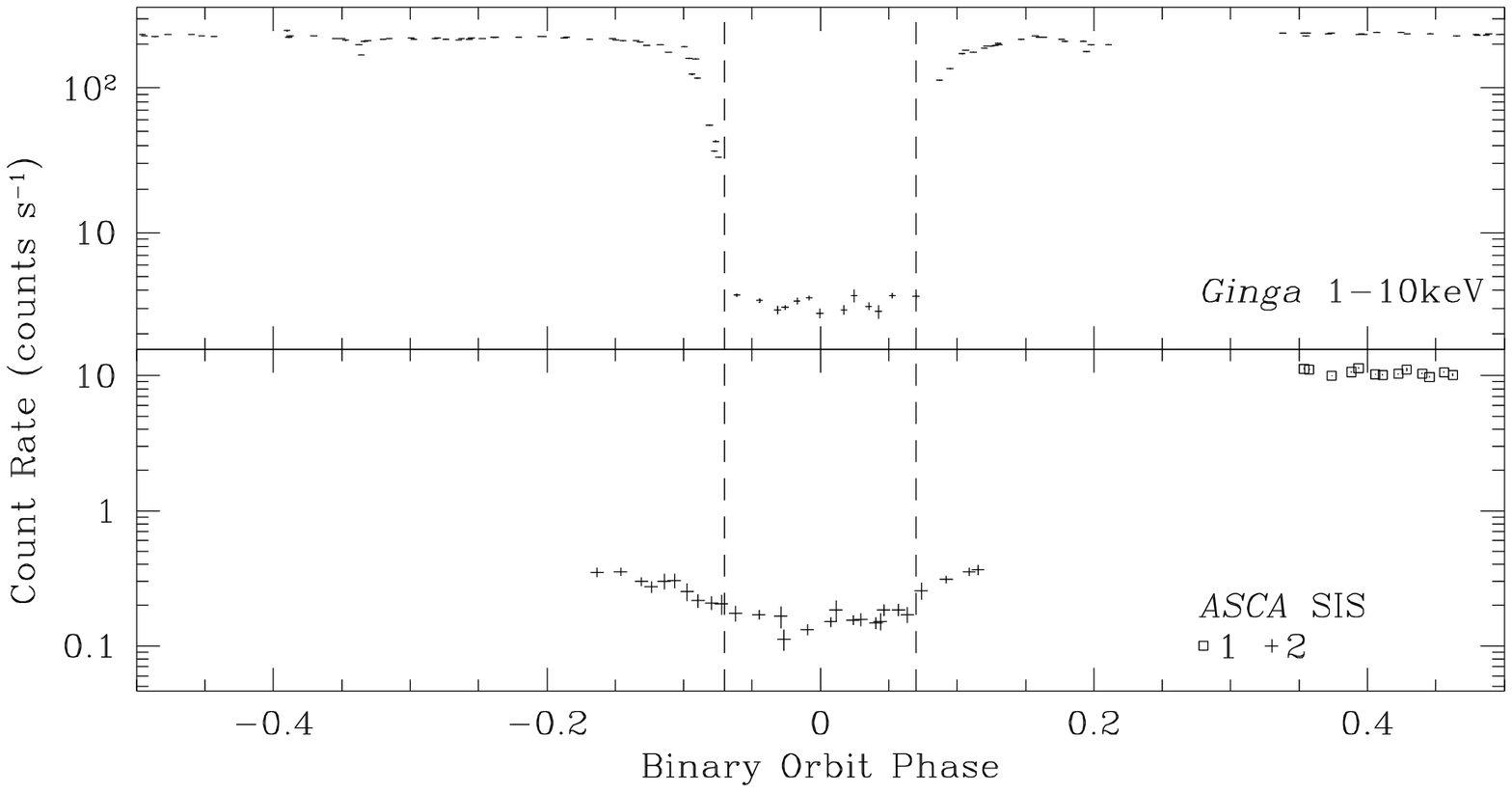]{Count rate of SMC~X-1 as a function of orbital
phase in the two observations (1 refers to the 1993 observation and 2
to the 1995 observation) with the {\it ASCA\/} SIS (count rate is the
sum of the two detectors) and in an observation with {\it Ginga}
(count rate is in the 1-10 keV band only). \label{lcs}}

\figcaption[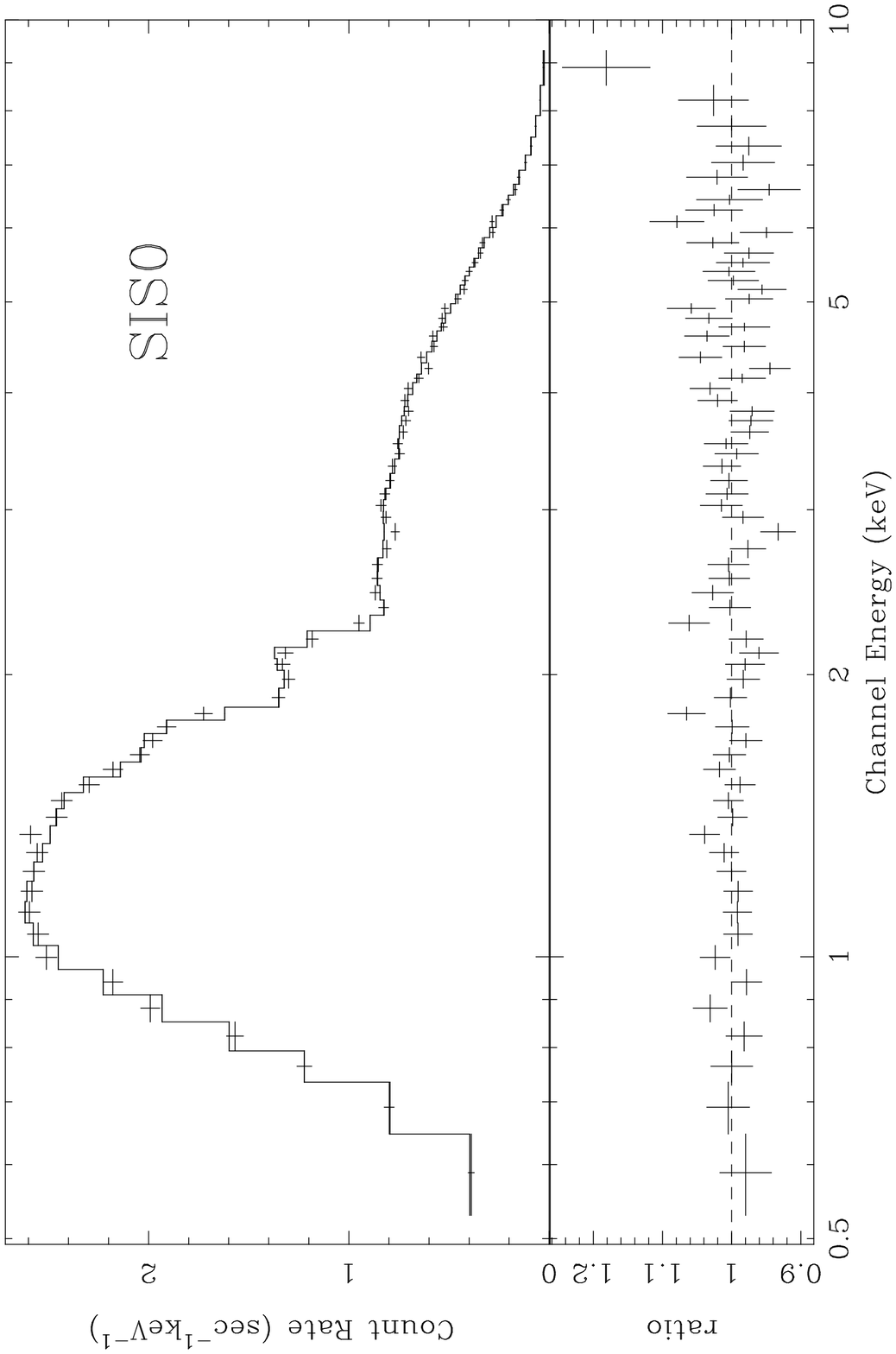,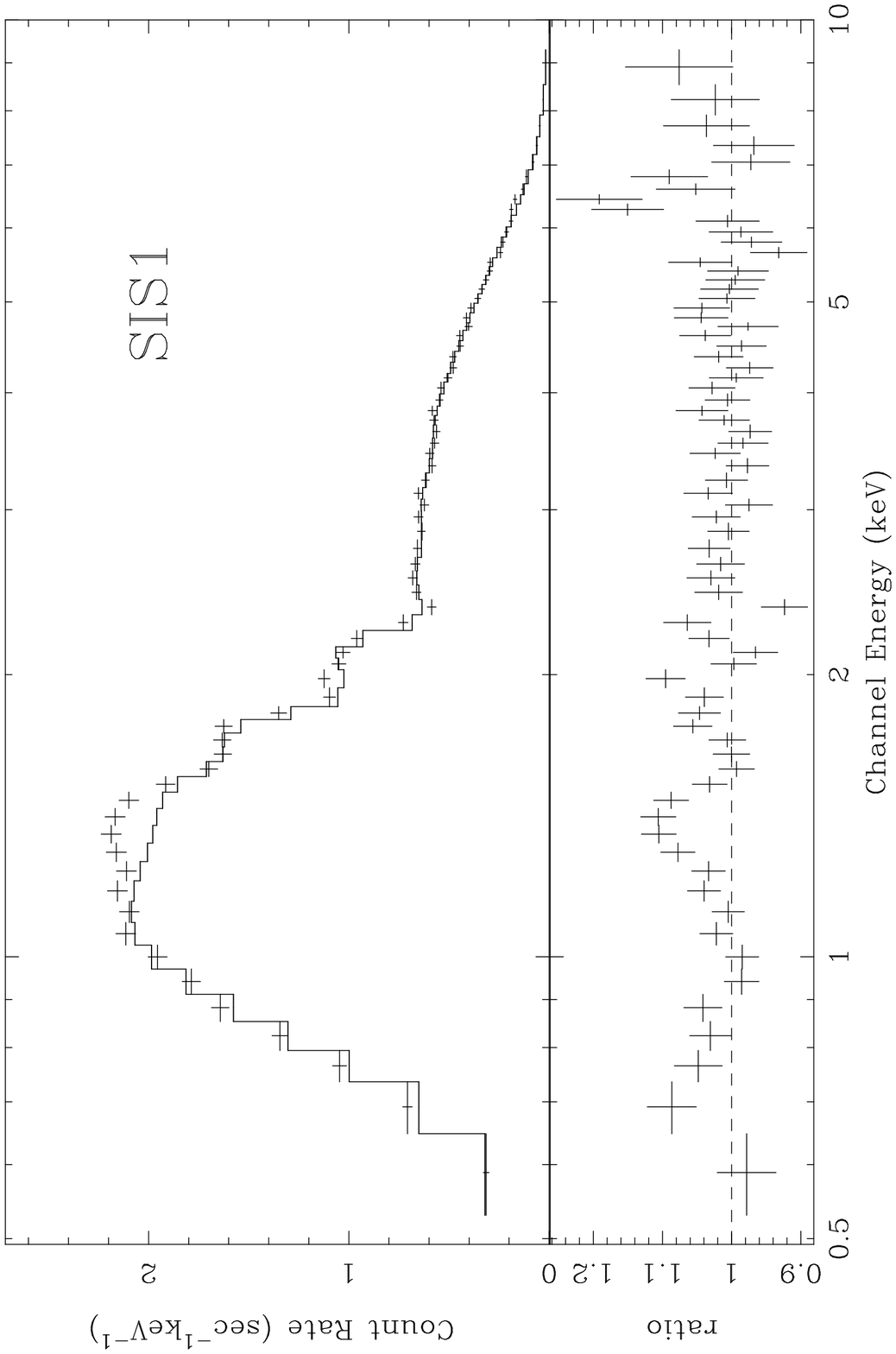]{Observed spectrum of SMC~X-1 out of
eclipse (crosses) and best-fit model (histogram). \label{unespec}}

\figcaption[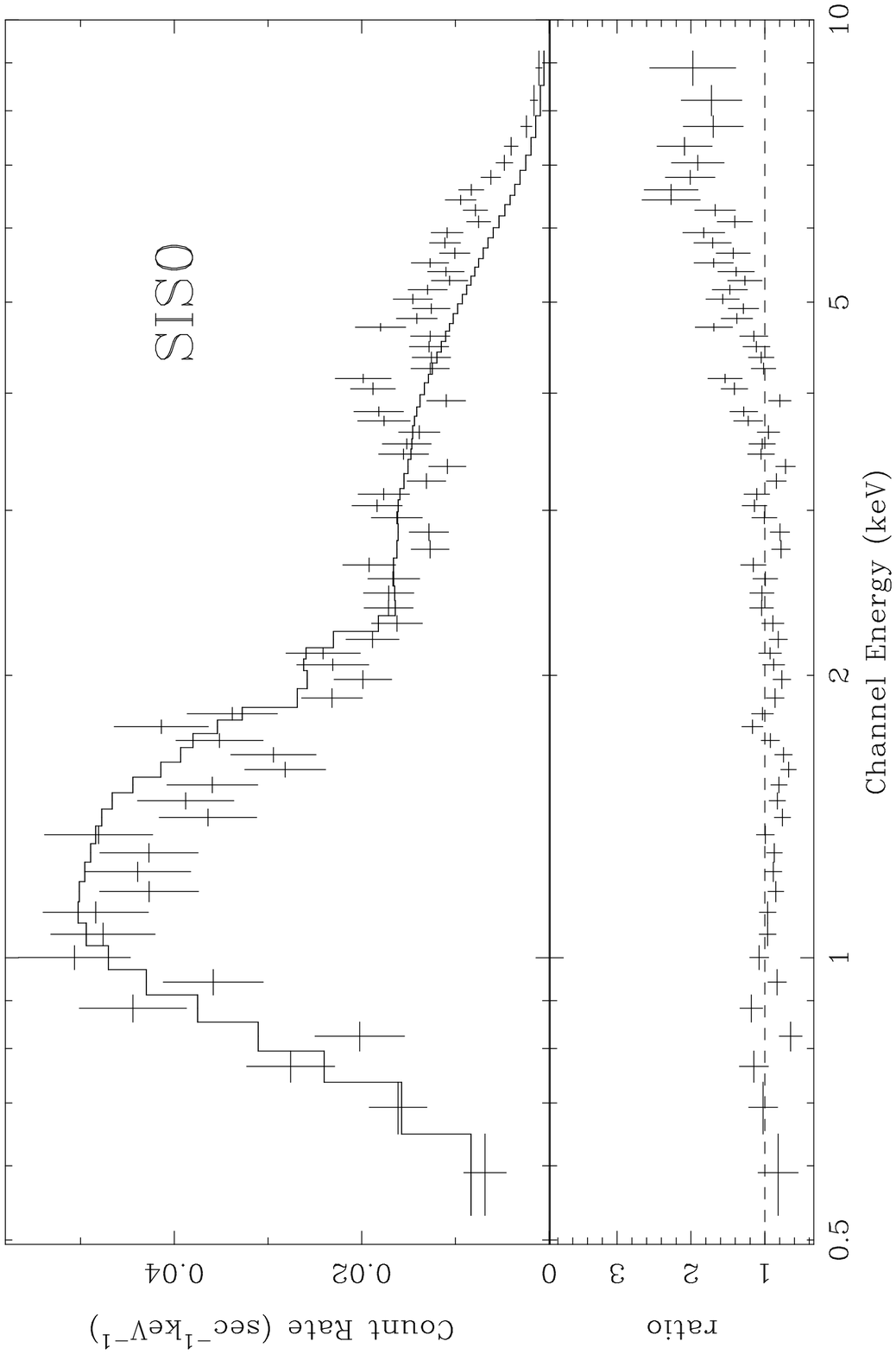,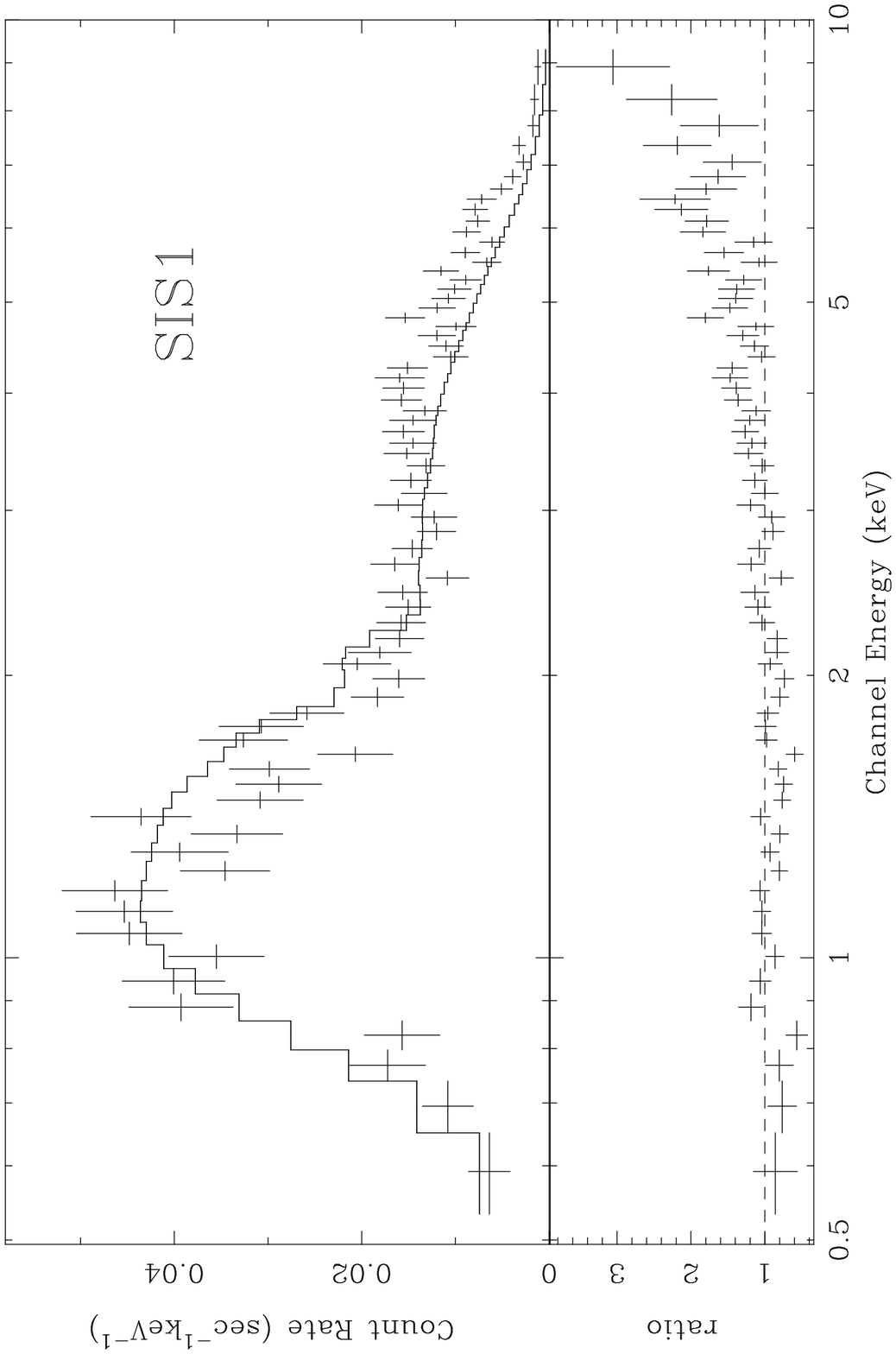]{Eclipse spectrum of SMC~X-1 with
out-of-eclipse model spectrum. \label{eclscal}}

\figcaption[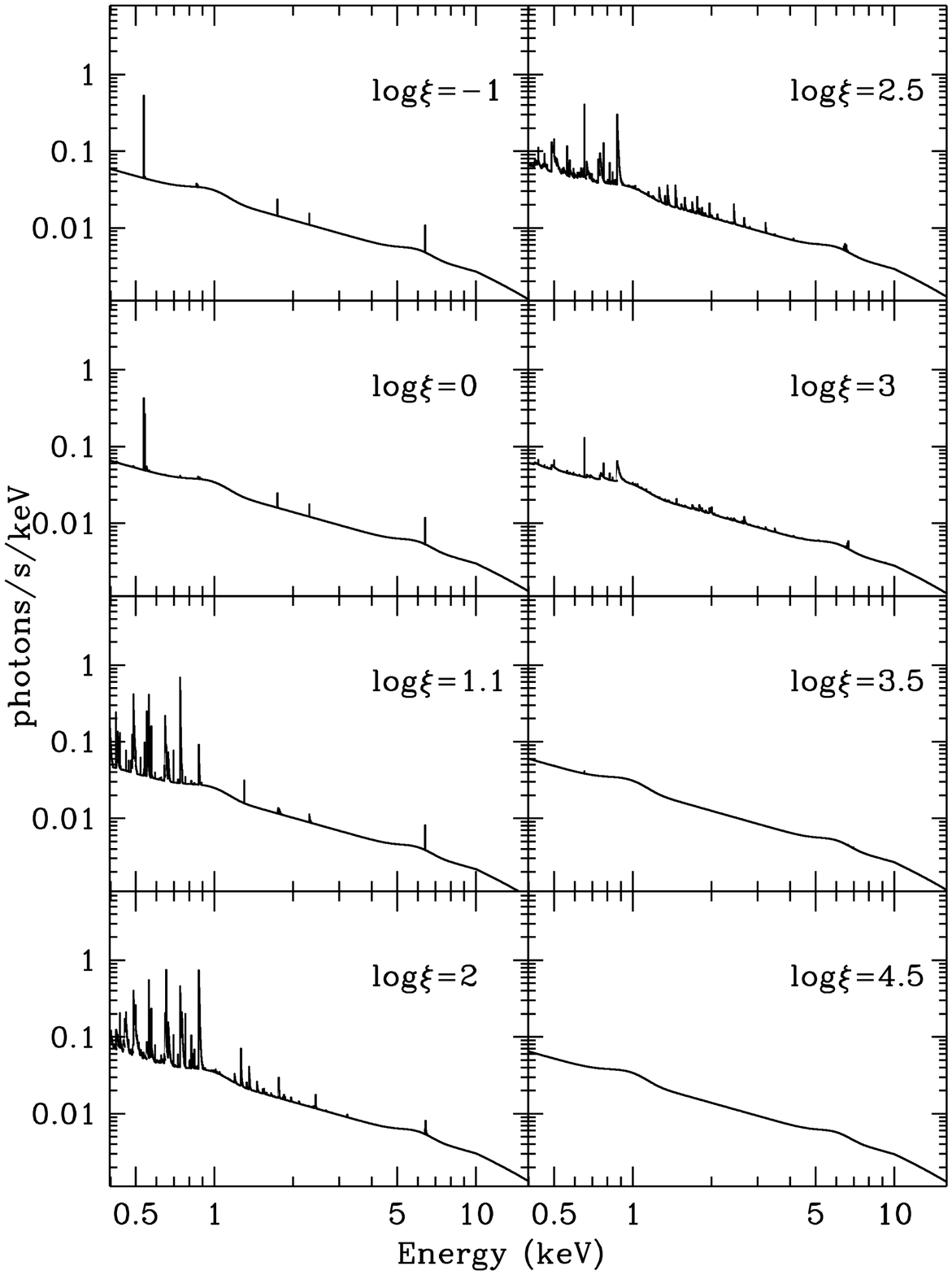]{Total spectra of reprocessed (Compton scattered,
bremsstrahlung, recombination, and fluorescent) radiation for gas when
illuminated by radiation of with the spectrum of the SMC~X-1
out-of-eclipse model for various values of $\log\xi$.  The
normalizations are arbitrary and chosen such that $n^2\xi$, which
determines the magnitude of the Compton scattered continuum is,
constant between the panels.  The resolution is 1340 bins per decade.
\label{spec_zeta}}

\figcaption[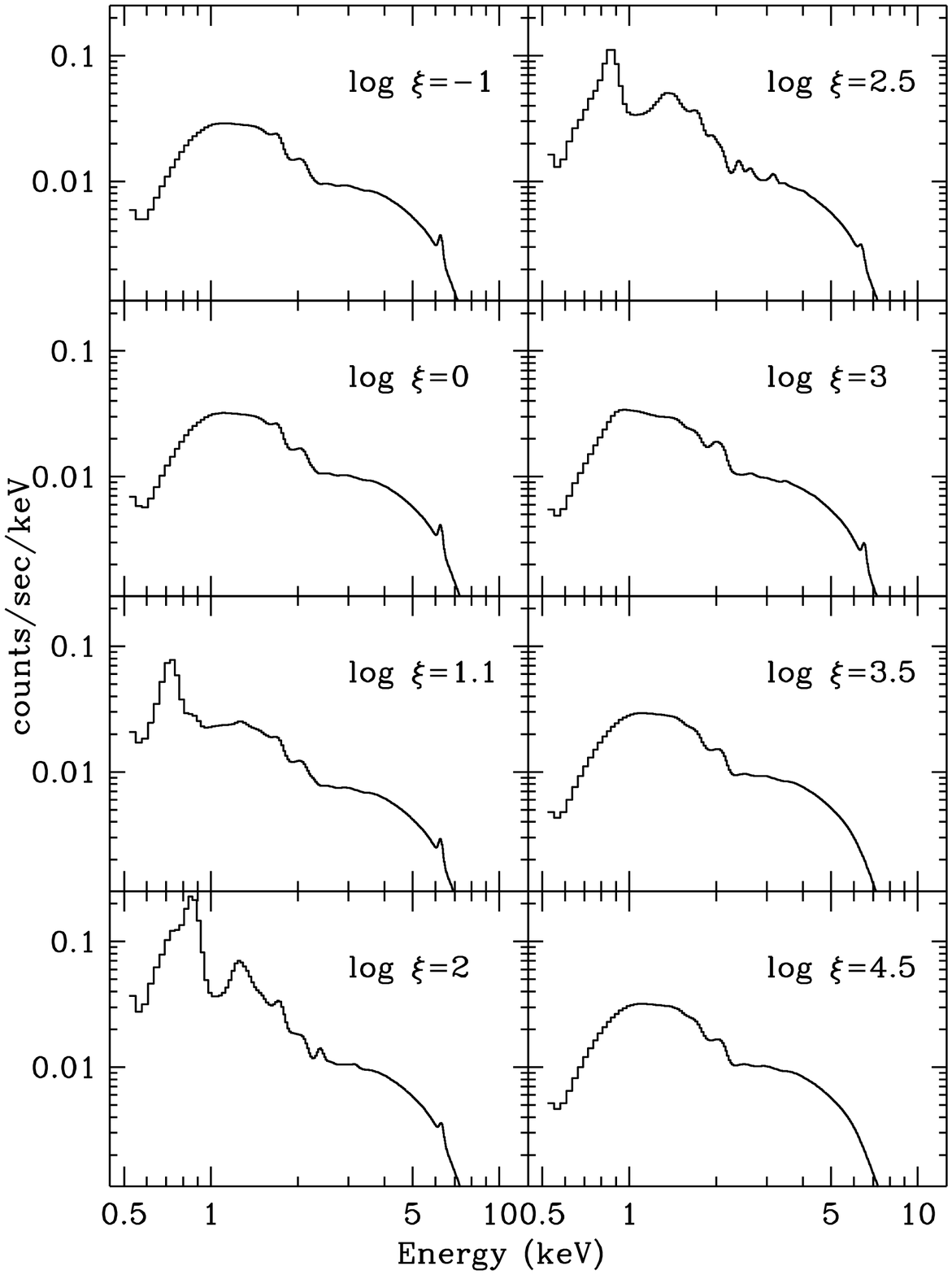]{The spectra of reprocessed radiation of
Figure~\ref{spec_zeta} absorbed by a column of $7\ee{20}$ and
convolved with the {\it ASCA\/} SIS0 response.
\label{folded_spec_zeta}}

\figcaption[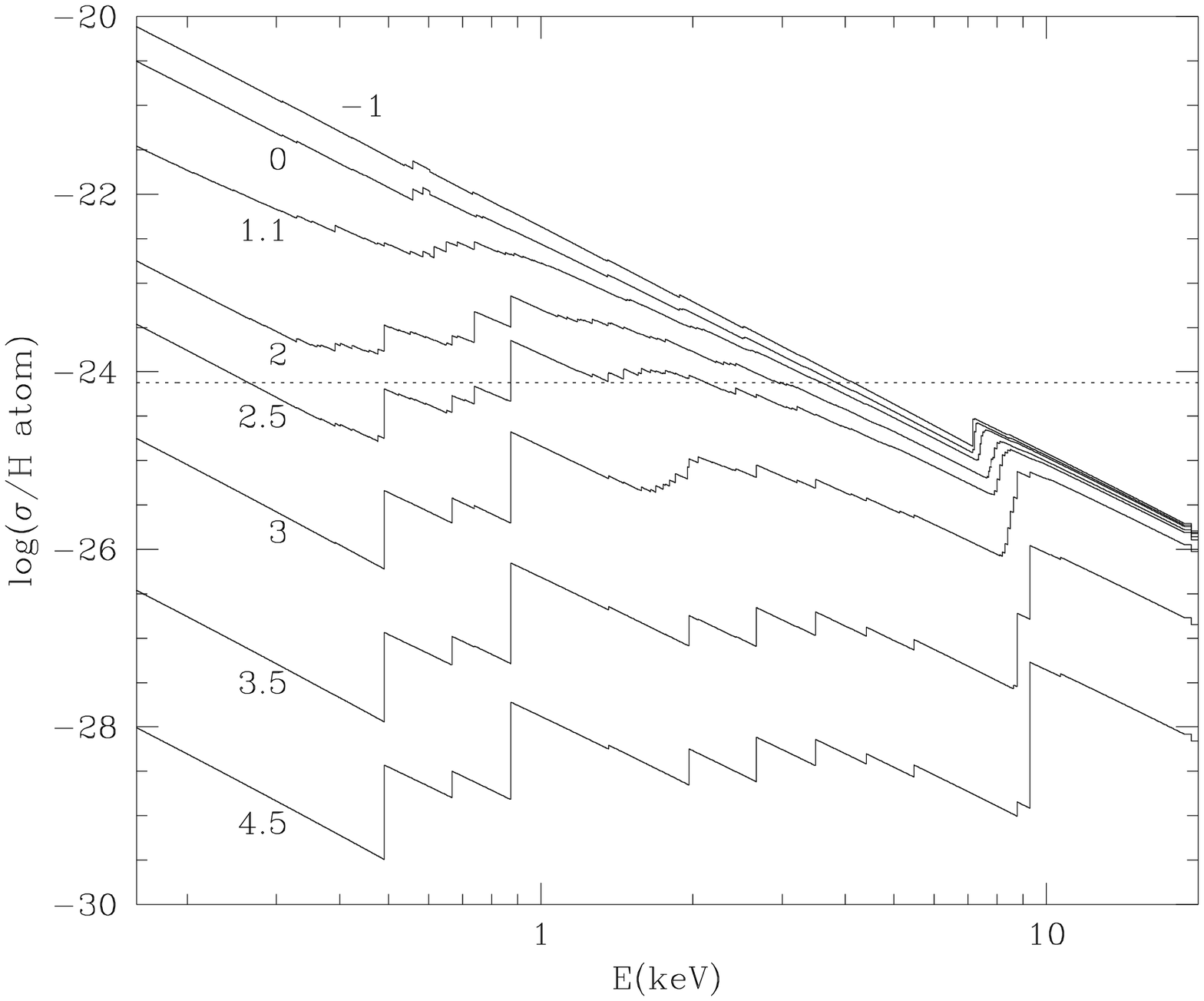]{Absorption cross-section per hydrogen atom for gas
at various values of $\log\xi$.  The dashed line is the Thompson
cross-section multiplied by 1.15 (the number of electrons per hydrogen
atom). \label{opac}}

\figcaption[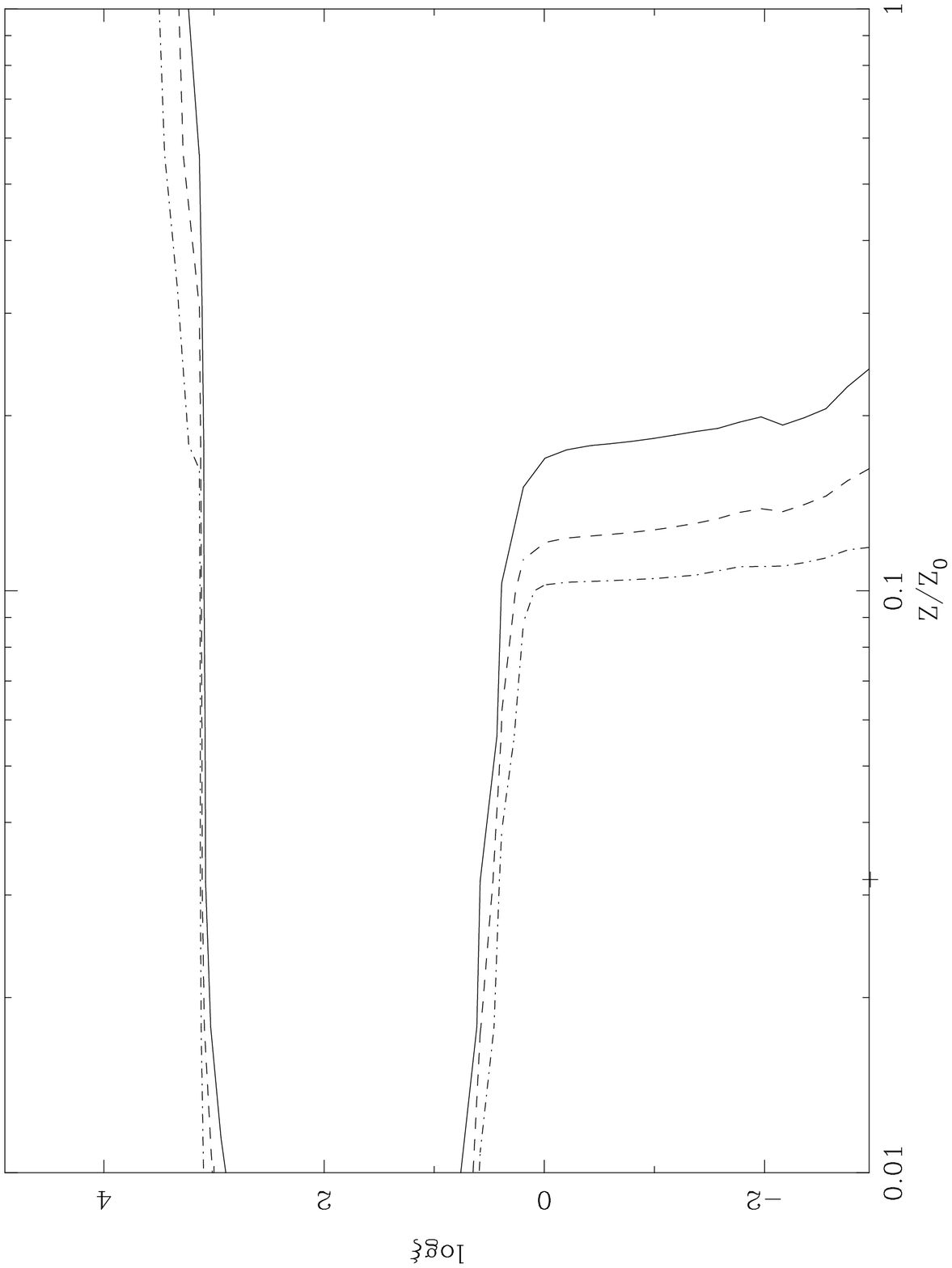]{Contours of $\chi^2$ for fits of the eclipse
spectrum to reprocessing by gas at single ionization parameters and
solar relative abundances.  The regions of good fits are in the top
and the lower left.  The contours mark $\Delta\chi^2=$2.3 (dot-dash),
4.61 (dash), and 9.21 (solid) from the best fit (marked with ``+'')
which has $\chi^2=112$ with 85 degrees of freedom. \label{cont_z_xi}}

\figcaption[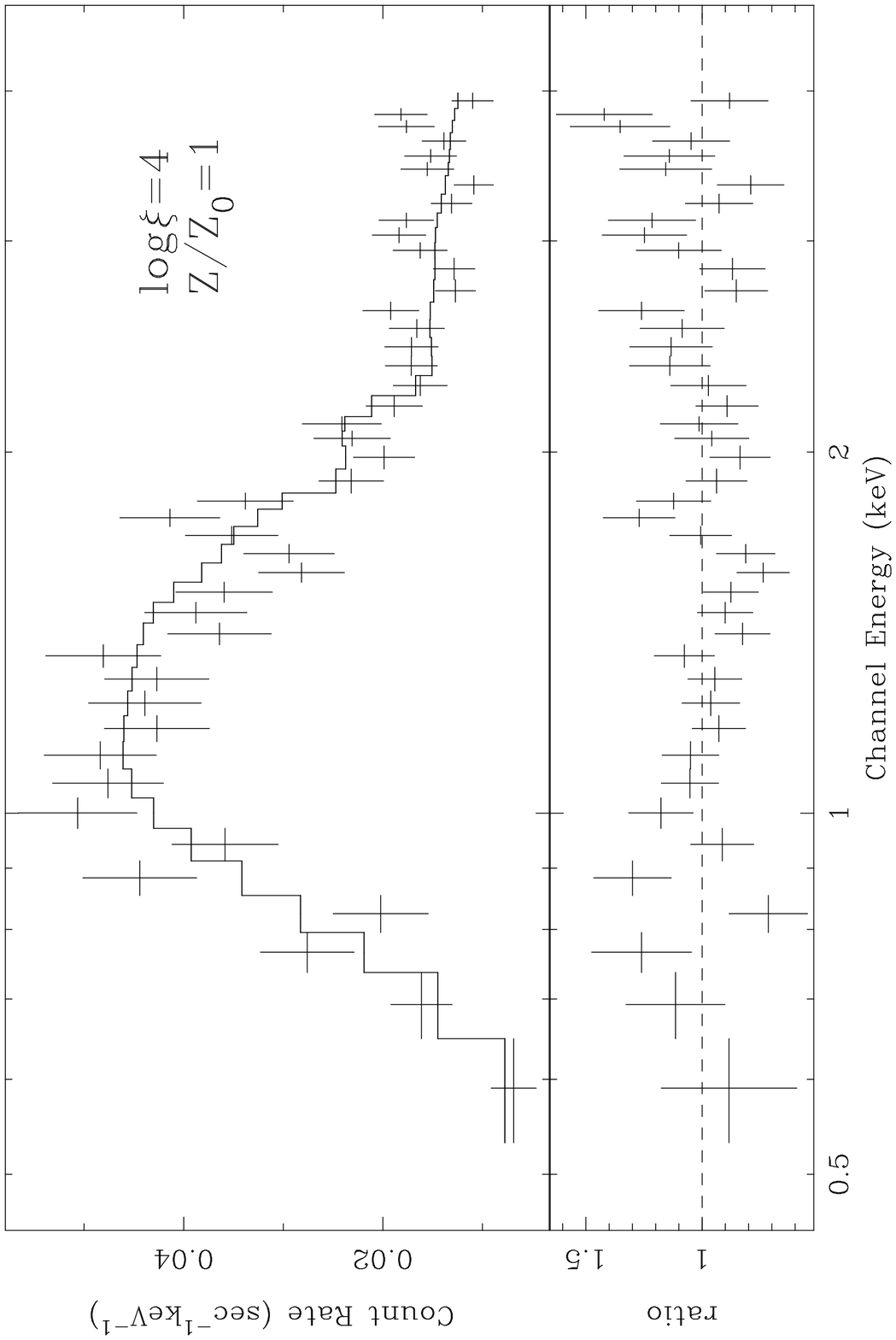,f8b_spec_zet18_ab-2_s0.eps,f8c_spec_zet00_ab-1_s0.eps]{SIS0
eclipse spectrum (crosses) compared to spectra of reprocessing from
gas at various single ionization parameters and solar relative
abundances. \label{spec_zeta_data_n_mo} }

\figcaption[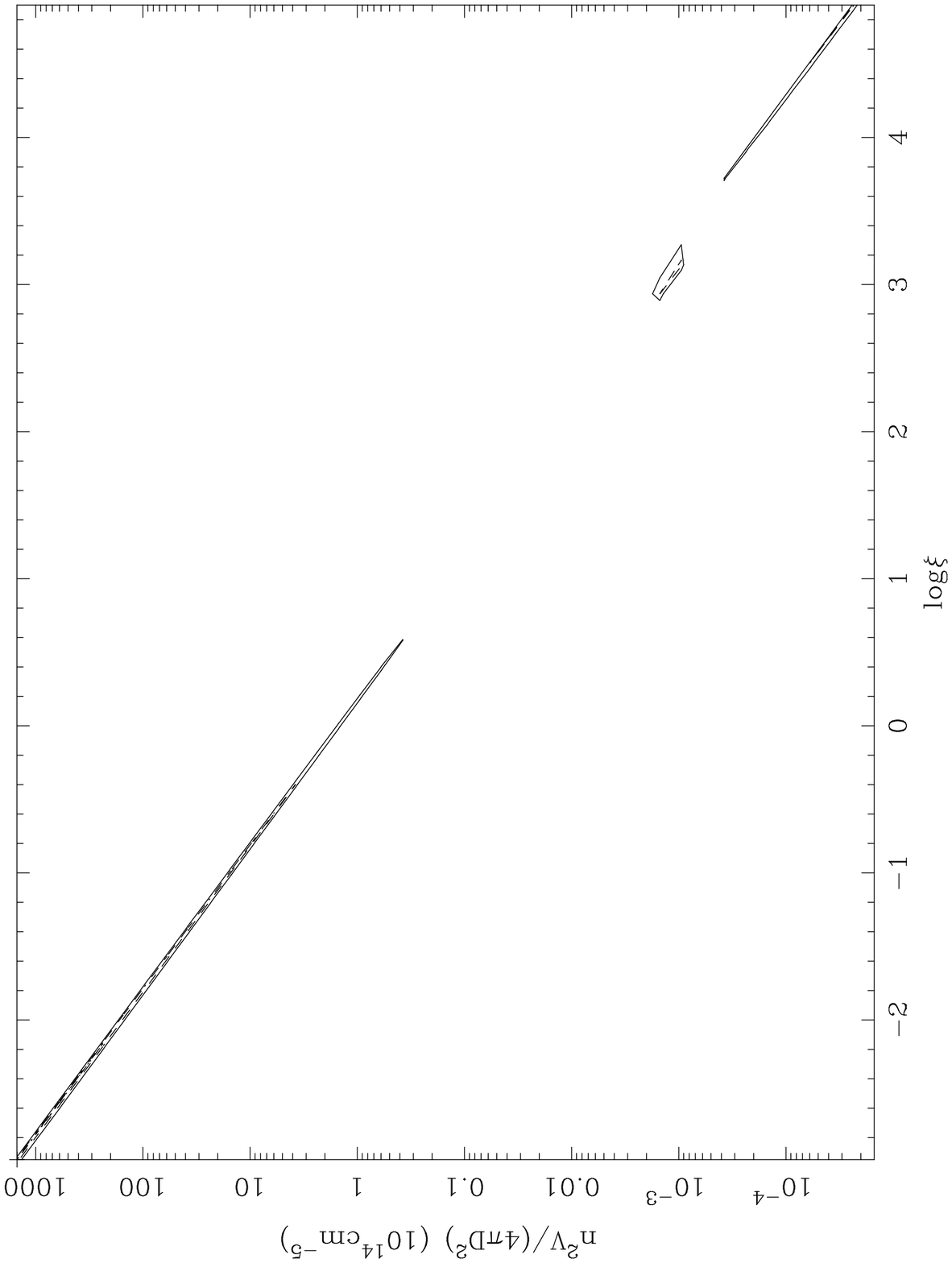]{The relation between $\xi$ and the normalization
$K$.  The solid contour marks $\Delta\chi^2=100$ relative to the best
fit (marked by ``+''). \label{cont_xi_norm}}

\figcaption[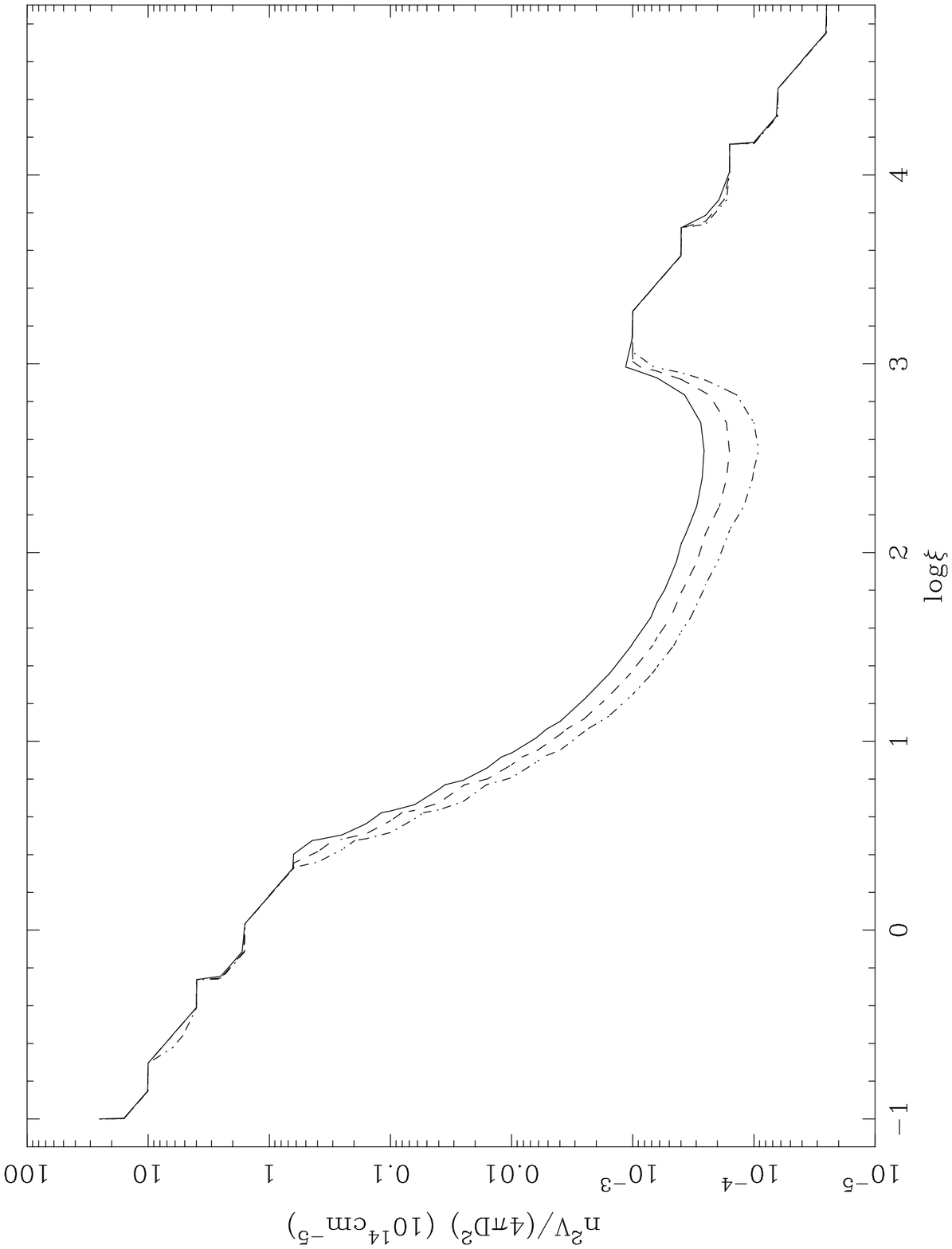]{Contours of $\Delta\chi^2$ for a $K$ and
$\log\xi$ of a second component. The contours mark
$\Delta\chi^2=$2.3(dot-dash), 4.6(dash), and 9.2(solid) relative to
the best fit (marked by ``+''). \label{cont_xi_norm_2comp}}

\figcaption[f11a.epsi,f11b.epsi,f11c.epsi,f11d.epsi]{Contours of $\log n$
(left panels) and $\log \xi$ (right panels) for the hydrodynamic
simulation of {\protect \citet{blo95}}.  The top panels are the
orbital plane and the bottom panels are the plane which contains the
axis of rotation and the line of centers.  The location of the neutron
star is marked with ``+''. \label{blo_pict} \notetoeditor{Placement
of figure panels: top left: a, bottom left: b, top right: c, bottom
right: d}}

\figcaption[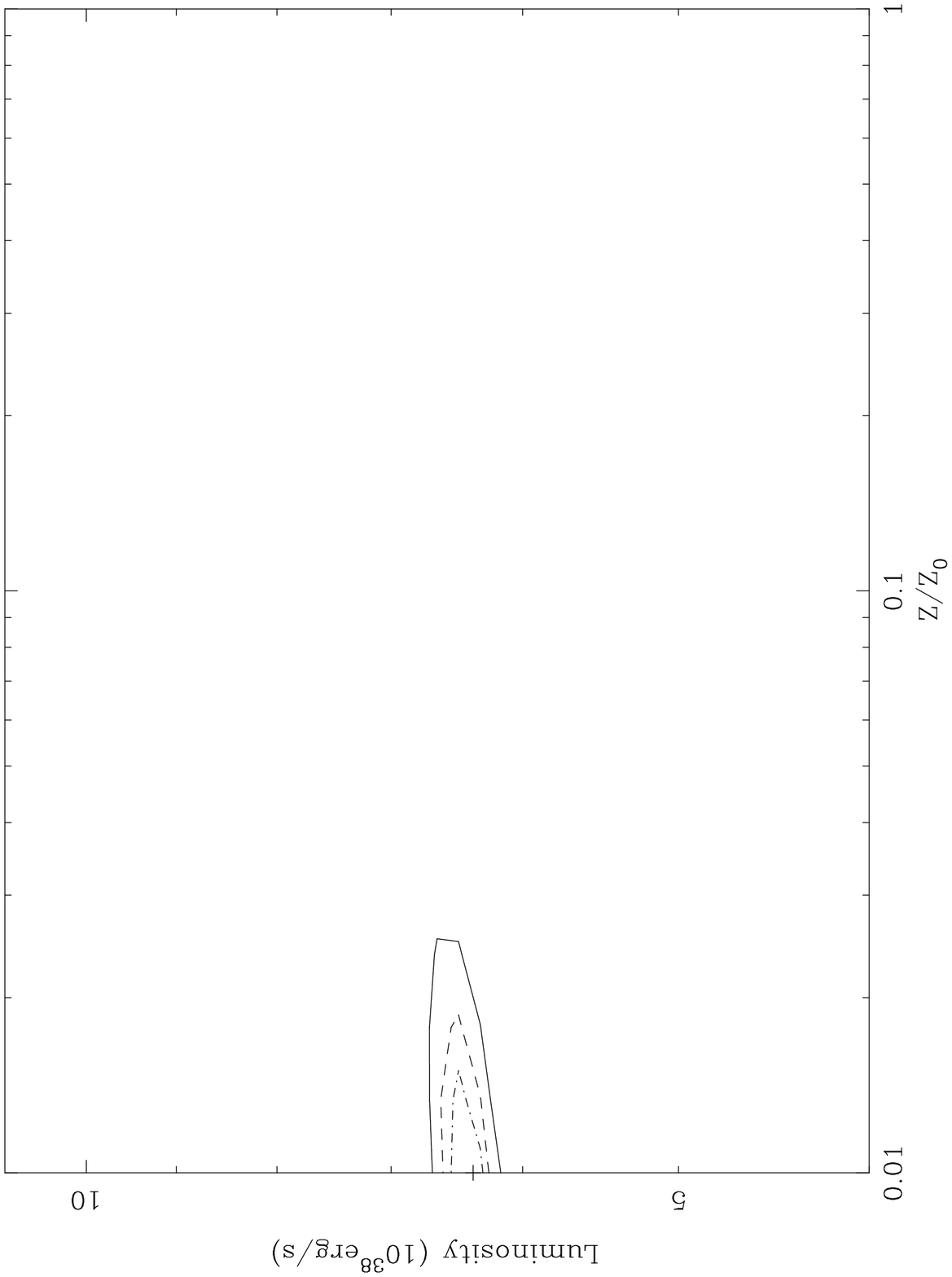]{Contours of $\chi^2$ for the spectral simulation
with the {\protect \citet{blo95}} matter distribution.  The contours
mark $\Delta\chi^2=$2.3 (dot-dash), 4.61 (dash), and 9.21 (solid) from
the best fit (marked with ``+'') which has $\chi^2=47.5$ with 41
degrees of freedom. \label{cont_z_l_blo_fixd}}

\figcaption[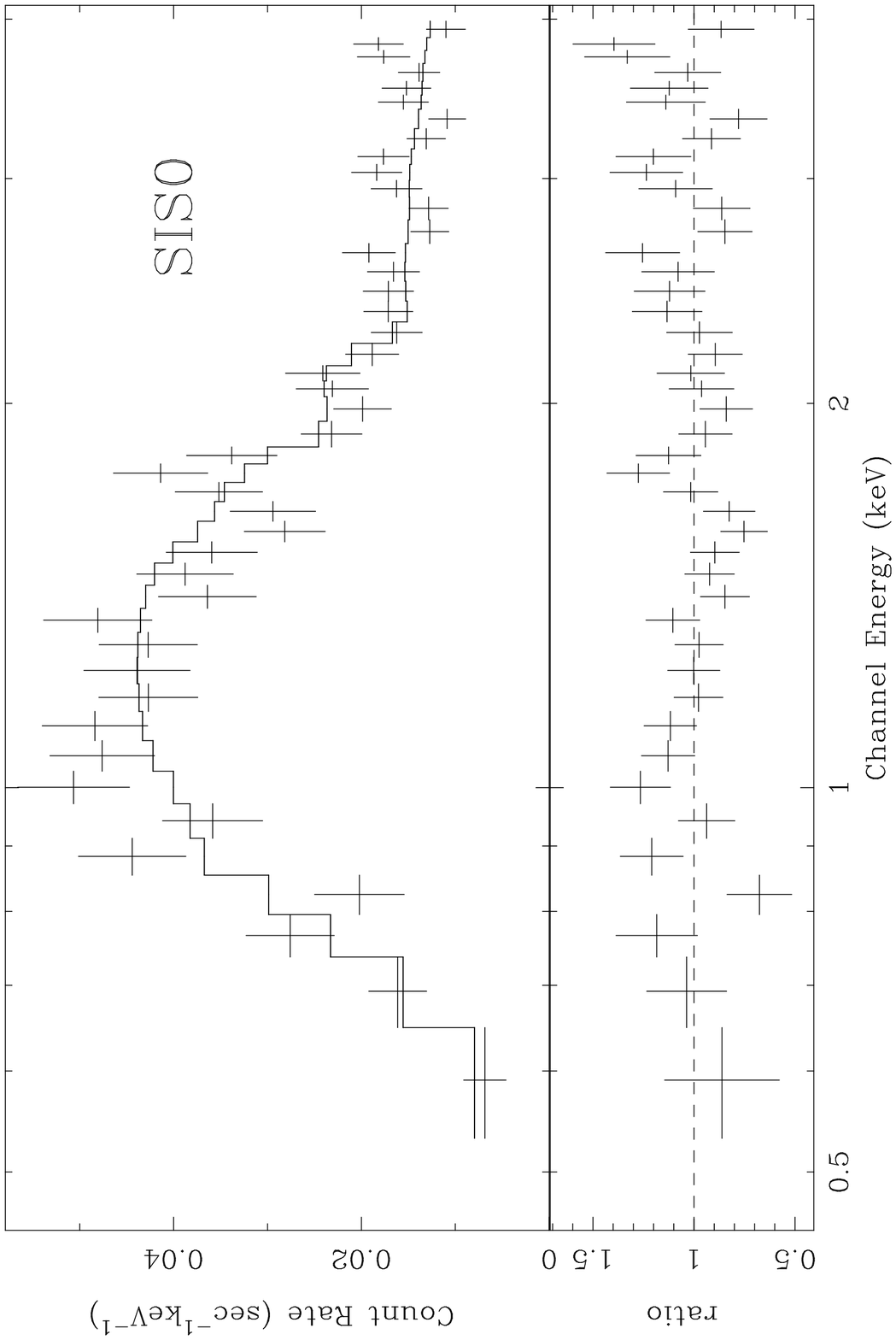]{SIS0 eclipse spectrum (crosses) and best fit
synthetic spectrum for the hydrodynamic simulation (histogram, $L_{\rm
X}=6.4\ee{38}$ erg~s$^{-1}$, $Z/Z_{\sun}=0.01$, $\chi^2=112$ for 86
degrees of freedom). \label{spec_blo_bf}}

\figcaption[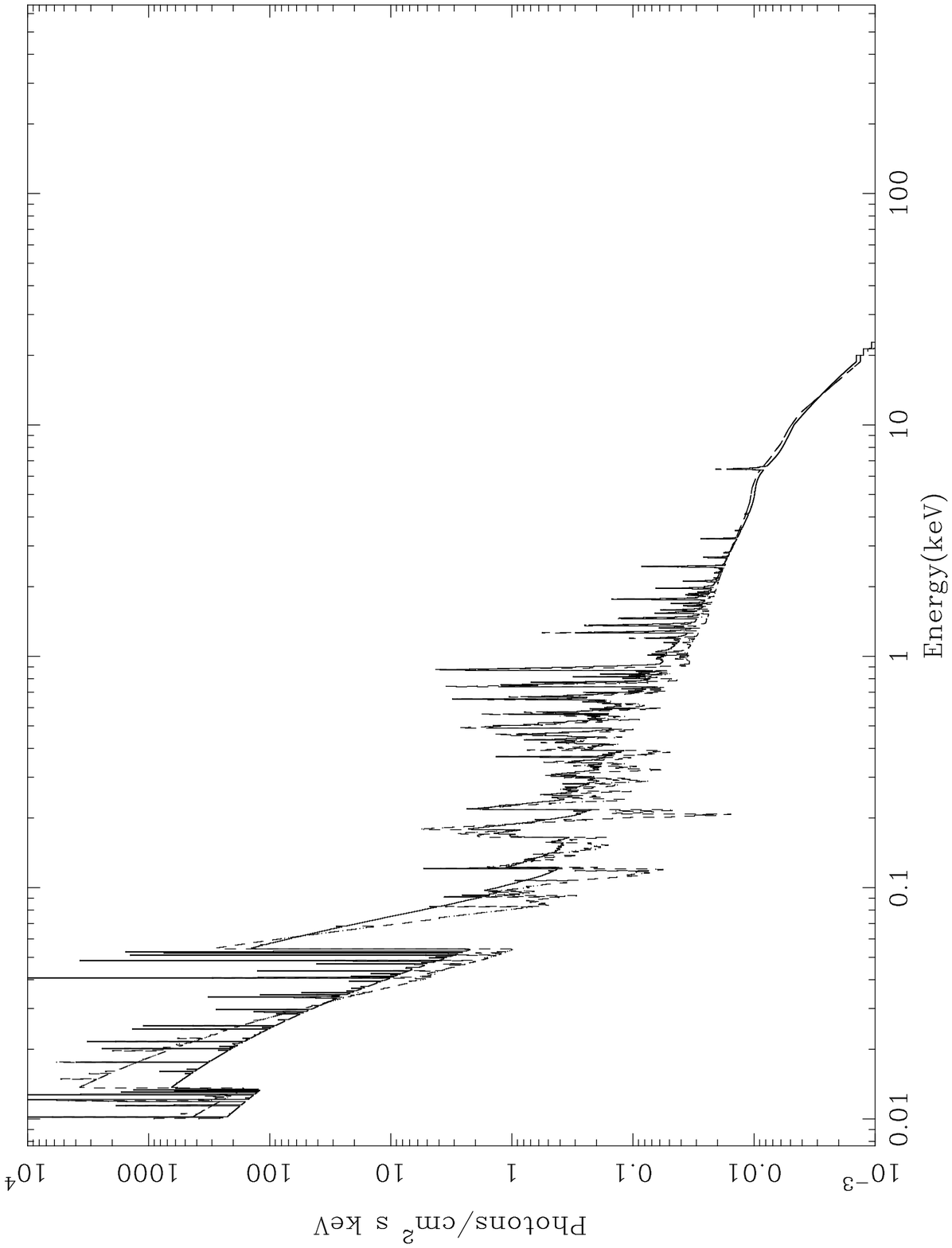,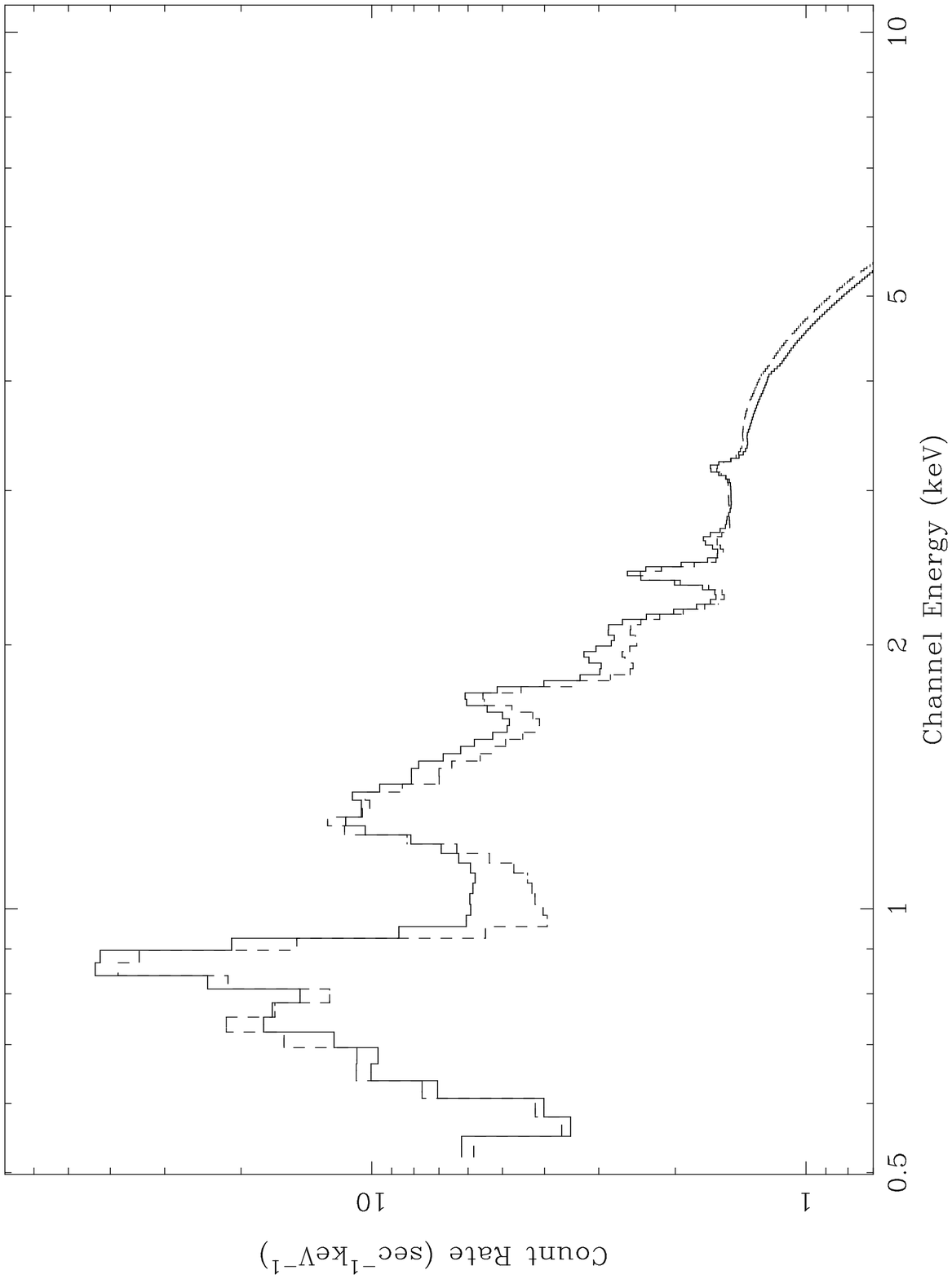,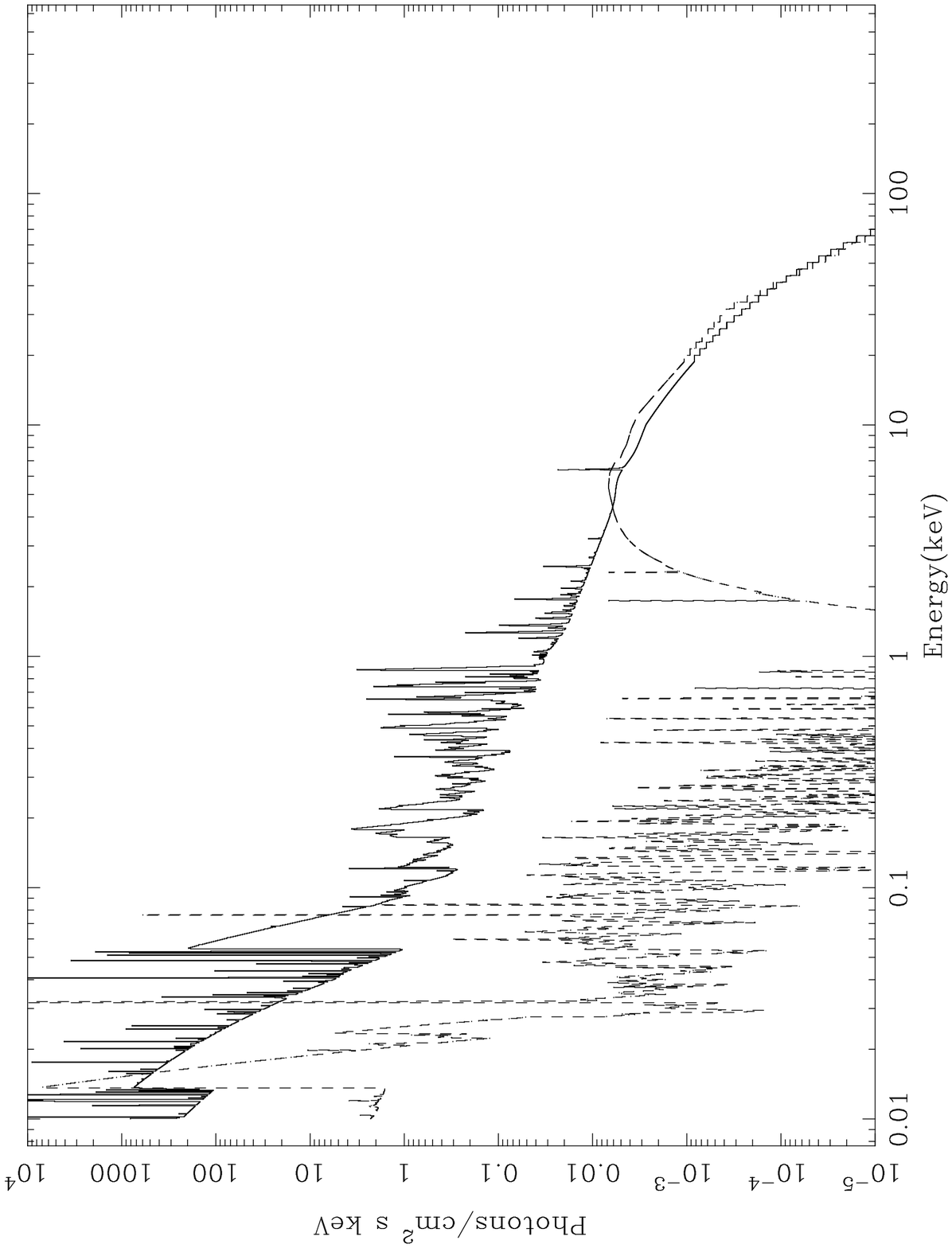,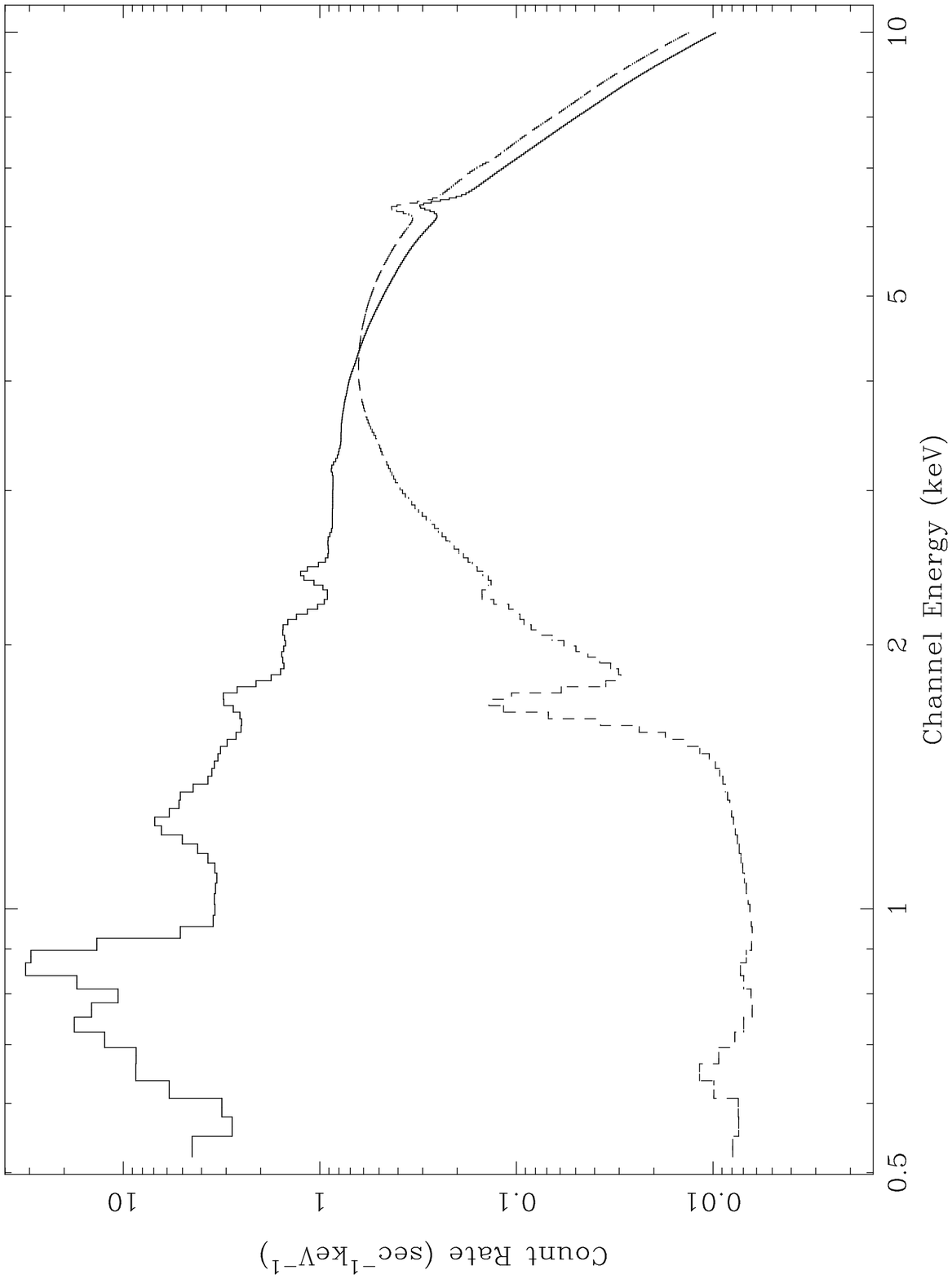]{Spectra of
reprocessed radiation in the optically thick case.  The solid lines
show the spectra of reprocessed radiation from optically thick gas.
In the left panels, $\log\xi=2.22$ and the primary radiation has
traveled through $10^{23}$ cm$^{-2}$.  In the right panels,
$\log\xi=1.98$ and the primary radiation has travelled through
$7.15\ee{23}$ cm$^{-2}$.  The top panels show the raw spectra and the
bottom panels show the spectra convolved with the {\it ASCA response}.
In all panels, the dashed lines indicate the spectrum of reprocessing
for the same ionization parameter in optically thin
material. \label{col}} \notetoeditor{Placement of figure panels: top
left: a, bottom left: b, top right: c, bottom right: d}

\figcaption[f15.epsi]{The differential emission measure of the {\protect
\citet{blo95}} wind distribution for $L=6\ee{38}$ erg
s$^{-1}$. \label{dem_blo}}

\figcaption[f16.epsi]{ Contours of
$\log\xi$ for a power-law wind plus exponential atmosphere.  The
ragged lines are an artifact. \label{hyb_zeta}} 

\begin{table}
\begin{center}
\begin{tabular}{l|r|rr} \tableline \tableline
                & \multicolumn{1}{|c|}{Uneclipsed} 
		& \multicolumn{2}{|c}{Eclipsed} \\
\hline
$n_{\rm H}$ (10$^{20}$cm$^{-2}$) & 7.0(1.0)       & 7.0(frozen)	& 7.0(frozen)  \\
$\alpha$ 	& 0.94(0.02)		& 0.94(frozen)	& 0.94(frozen) \\
$K_{\rm pl}$	& $2.67(0.07)\ee{-2}$	& $4.9(0.1)\ee{-4}$ & $3.77(0.20)\ee{-4}$ \\
$E_1$(keV)	& 0.93(0.04) 		& 0.93(frozen)	& 0.93(frozen) \\
$\sigma_1$	& 0.16(0.03)		& 0.16(frozen)  & 0.16(frozen) \\
$K_{\rm ga1}$	& $3.4(0.9)\ee{-3}$	& $6.1\ee{-5}$ 
(tied to $K_{\rm pl}$) & $4.3\ee{-5}$ (tied to $K_{\rm pl}$) \\
$E_2$		& 6.0(0.1)		& 6.0(frozen)	& 6.8(0.3) \\
$\sigma_2$	& 1.0(0.2)		& 1.0(frozen)  & 2.0(0.3)   \\
$K_{\rm ga2}$	& $2.2(0.8)\ee{-3}$	& $3.9\ee{-5}$
(tied to $K_{\rm pl}$) & $5.1(0.7)\ee{-4}$ \\
\tableline
$\chi^2$/d.o.f.	& 70/62			& 376/141 	& 139/138 \\
probability	& 21\%			& $<10^{-32}$  	& 46\%  \\
\end{tabular}
\caption[Spectral Fit Parameters]{Best fit spectral parameters and
1-sigma errors.}
\label{specpars}
\end{center}
\end{table}

\begin{deluxetable}{lrr}
\tablecolumns{8}  
\tablewidth{0pc}  
\tablecaption{Parameters of Reflection model fits}
\label{reffit}
\tablehead{
\colhead{} & \colhead{Uneclipsed} & \colhead{Eclipsed} }
\startdata
$n_{\rm H}$(10$^{20}$cm$^{-2}$) & 8.7(0.7) & 8.7(frozen) \\
$\alpha$ 	& 1.03(0.03)		&  1.03(frozen)	\\
$K_{\rm pl}$ ($10^{-2}$ ph s$^{-1}$ keV$^{-1}$ at 1 keV) & 2.81(0.06)
& 2.81(frozen) \\
$\sigma_1$ (keV)	& 0.16(0.02)		& 0.16(frozen) \\
$K_{\rm ga1}$ ($10^{-3}$ph s$^{-1}$) 	& 3.2($^{+1.0}_{-0.5}$) &
3.2(frozen) \\ 
Escape Fraction & 1(frozen)	& 1.38(0.03)$\ee{-2}$ \\
Covering Fraction & 2.9(0.15)	& 0.22(0.01) \\
\tableline
$\chi^2$/d.o.f.	& 80/64		&  147/40 \\
probability	& 9\%		&  33\%  \\
\enddata
\end{deluxetable}


\clearpage
\plotone{f1.ps}

\epsscale{0.4}
\clearpage
\plotone{f2a.eps}
\plotone{f2b.eps}
 
\clearpage
\plotone{f3a.eps}
\plotone{f3b.eps} 

\epsscale{1.0}
\clearpage
\plotone{f4.ps}

\clearpage
\plotone{f5.ps}

\clearpage
\plotone{f6.eps}

\plotone{f7.eps}

\epsscale{0.3}
\clearpage
\plotone{f8a.eps}
\plotone{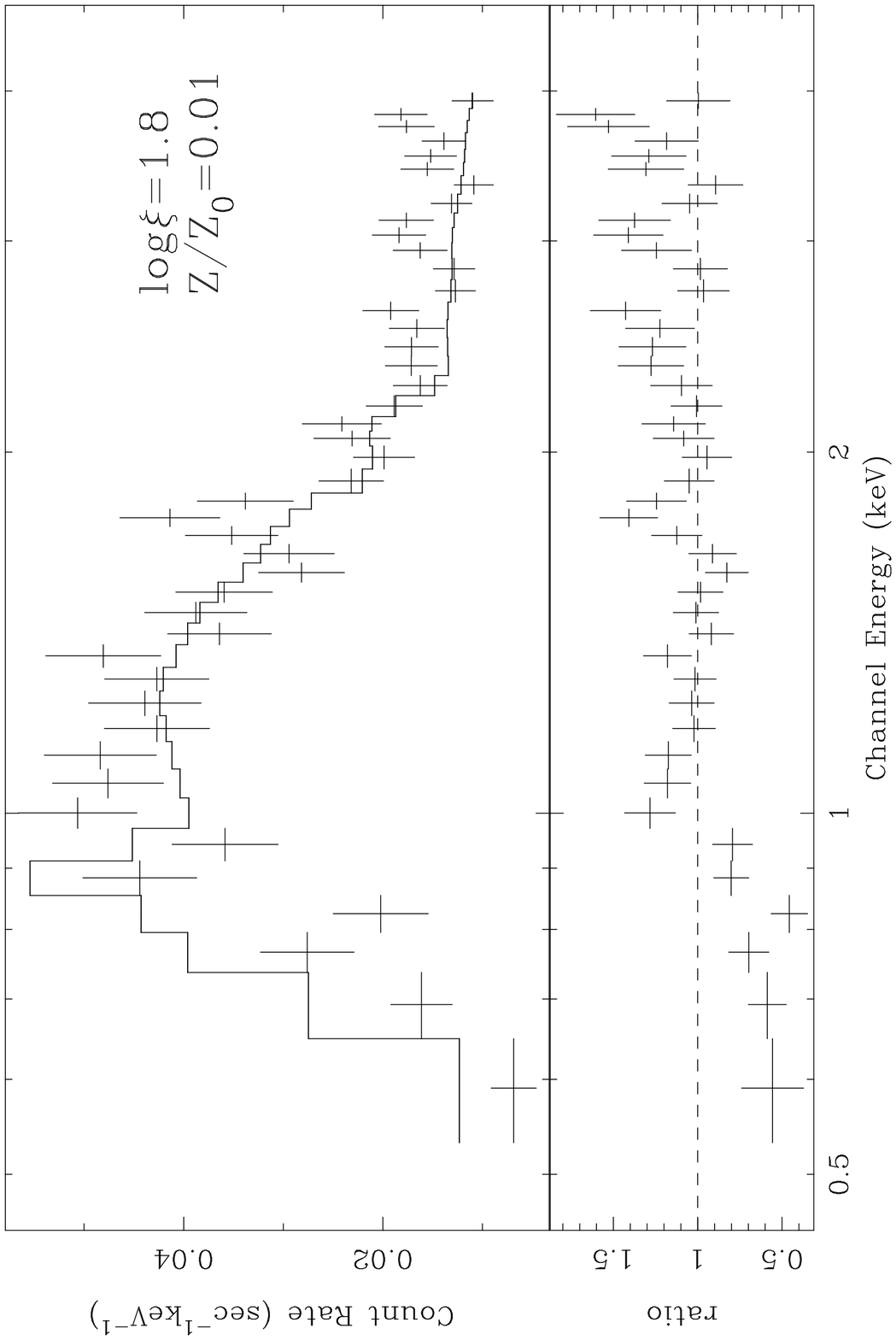}
\plotone{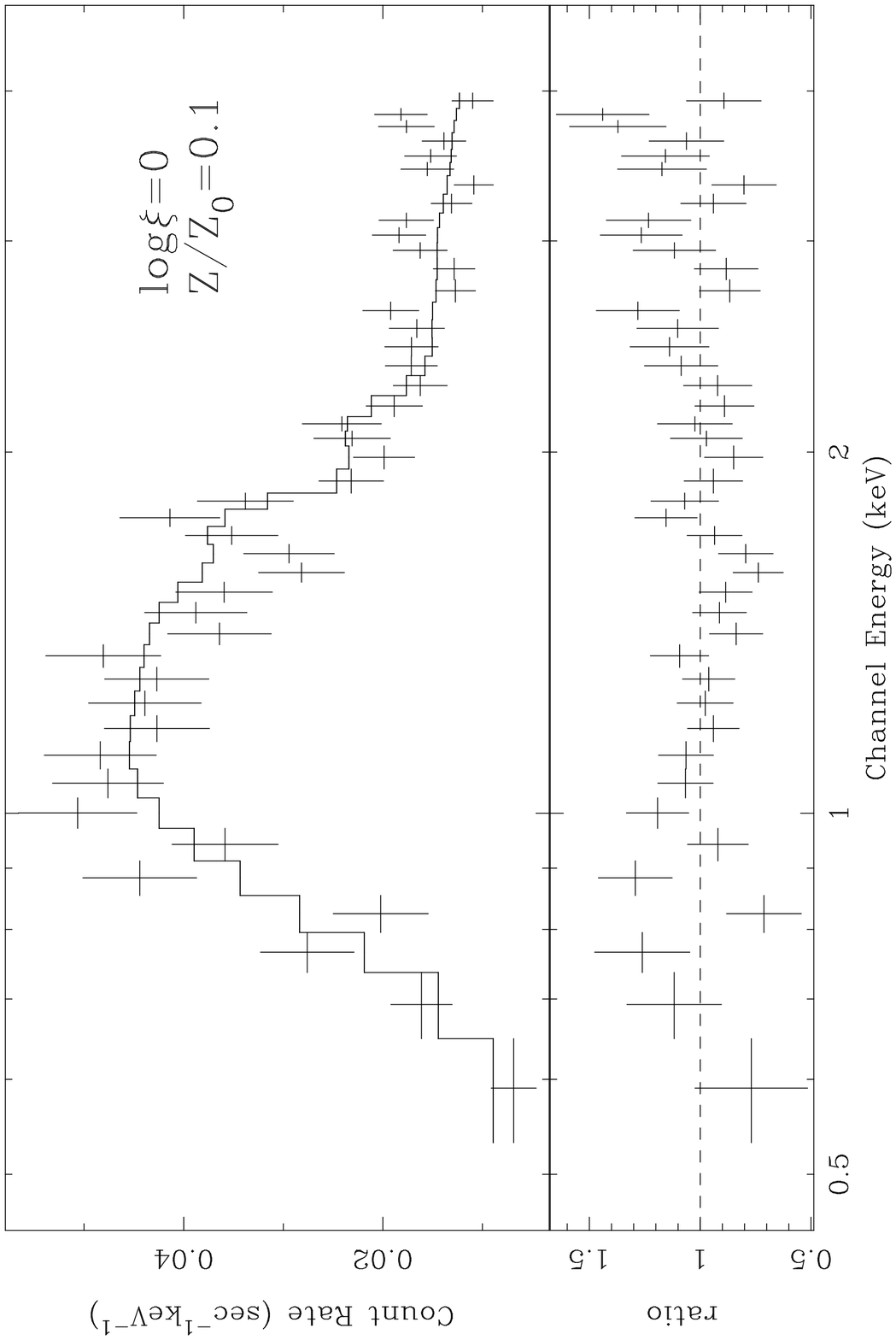}

\epsscale{}
\plotone{f9.eps}

\plotone{f10.eps}

\clearpage
\epsscale{0.4}
\plotone{f11a.epsi}\plotone{f11c.epsi} 
\epsscale{0.4}
\plotone{f11b.epsi}\plotone{f11d.epsi}

\clearpage
\epsscale{0.8}
\plotone{f12.eps}

\clearpage
\plotone{f13.eps}

\clearpage
\epsscale{0.4}
\plotone{f14c.eps}\plotone{f14d.eps} \\
\plotone{f14a.eps}\plotone{f14b.eps}

\epsscale{1.0}
\clearpage
\plotone{f15.epsi}

\epsscale{1.0}
\clearpage
\plotone{f16.epsi}

\end{document}